\documentclass[useAMS]{mn2e}

\usepackage{psfig}

\newcommand{\mdot}{\rm{M_{\odot}}}
\newcommand{\bh}{\rm{\scriptscriptstyle{BH}}}

\title[Short time-scale optical variability in NGC 4395]
{Short time-scale optical variability of the dwarf Seyfert nucleus in
NGC 4395}

\author[J. E. Skelton et al.]{J. E. Skelton$^{1}$, A. Lawrence$^{1}$, A. Pappa$^{1}$,
P. Lira$^{2}$, O. Almaini$^{1,3}$\\
$^{1}$Institute for Astronomy, University of Edinburgh, Royal
Observatory, Blackford Hill EH9 3HJ\\
$^{2}$Departamento de Astronom\'{i}ca, Universidad de Chile, Casilla 36-D,
Santiago, Chile\\
$^{3}$School of Physics and Astronomy, University of Nottingham,
University Park, Nottingham NG7 2RD\\}

\pagerange{\pageref{firstpage}--\pageref{lastpage}} \pubyear{2004}

\begin{document}

\maketitle

\label{firstpage}

\begin{abstract}
We present optical spectroscopic observations of the least-luminous
known Seyfert 1 galaxy, NGC 4395, which was monitored every half-hour
over the course of 3 nights.  The continuum emission varied by
$\sim$35\ per cent over the course of 3 nights, and we find marginal evidence
for greater variability in the blue continuum than the red.  A number of diagnostic checks
were performed on the data in order to constrain any systematic or
aperture effects.  No correlations were found that adequately
explained the observed variability, hence we conclude that we have
observed real intrinsic variability of the nuclear source. No
simultaneous variability was measured in the broad H$\beta$ line,
although given the difficulty in deblending the broad and narrow
components it is difficult to comment on the significance of this result.   The
observed short time-scale continuum variability is consistent with NGC 4395
having an intermediate-mass ($\sim 10^5\rm{M_{\odot}}$) central supermassive black hole,
rather than a very low accretion rate.  Comparison with the Seyfert 1
galaxy NGC 5548 shows that the observed variability seems to scale with
black hole mass in roughly the manner expected in accretion models. However the
absolute timescale of variability differs by several orders of magnitude from
that expected in simple accretion disc models in both cases.
\end{abstract}


\section{Introduction}

NGC 4395 is a nearby, late-type spiral galaxy of morphological type Sd
III-IV.  It exhibits a very low surface-brightness disk, with almost
no central bulge and loose spiral arms showing some isolated regions
of star formation.  Optical spectra of the faint, star-like nucleus
reveal strong, narrow forbidden emission lines from a wide range of
species and ionisation, as well as weak broad wings on permitted
lines.  Filippenko \& Sargent (1989) interpreted these observations as an indication
that NGC 4395 harboured an extremely low-luminosity Seyfert 1
nucleus.  Subsequent multi-wavelength observations (Sramek et al. 1992;
Filippenko, Ho \& Sargent 1993) confirmed this classification and
convincingly argued against other origins for the observed non-stellar
emission.  With
an absolute blue magnitude of $\sim -11$, the nucleus of NGC 4395 has a
lower luminosity than the brightest supergiant stars, and is a factor
of $\sim 10^4$ fainter than even low-luminosity classical Seyfert 1s such as NGC 4051,
making it the least-luminous Seyfert 1 discovered to date.

There are two likely possibilities for the very low luminosity nucleus in
NGC 4395.  One is a low accretion rate, in line with other
low-luminosity AGN (Ptak et al. 1998).  The other is that the central
supermassive black hole is significantly smaller than in more luminous
AGN, with a relative accretion rate that is more typical of classical
Seyferts giving rise to the observed properties of NGC 4395.  Given
that NGC 4395 is a small late-type spiral with a very low-luminosity
bulge component, a small
central black hole is consistent with the observed correlation between
bulge mass and black hole mass (Ferrarese \& Merritt 2000; Gebhardt et al. 2000).  Several
estimates of the black hole mass in NGC 4395 have been made.  A
firm upper limit of $6.2 \times 10^6\rm{M_{\odot}}$ has been placed by
Filippenko \& Ho (2003) using velocity dispersion analysis of the
Ca {\sc ii} near-IR triplet.  However this is a measure of the virial
mass of the entire nuclear stellar cluster and the mass of the actual
black hole could be considerably smaller.  Estimates using
photoionisation modelling of the broad-line region (Kraemer et al. 1999) and from power-law fitting to the fluctuation power density
spectrum (Shih et al. 2003) give
consistent estimates in the region $10^4 - 10^5\rm{M_{\odot}}$.  These
values are lower by several orders of magnitude than those found in
other typical Seyfert galaxies, and while neither of these methods is
particularly accurate, it is encouraging that the estimates are
consistent with one another.   

One test of the size of the central engine in NGC 4395 is fast
multi-wavelength variability.  Smaller scale-lengths imply that the
light- and sound-crossing time-scales should be shorter, leading to
variability on short time-scales.  An early optical
monitoring campaign over several years by Shields \& Filippenko
(1992) reported that NGC 4395 showed no variability in either the
broad lines or the continuum.  However, a subsequent spectroscopic study by
Lira et al. (1999)  found continuum variability over a time-scale
of 6 months of factors of order 2 and 1.3 in the blue and red
respectively.  Variability in the broad lines was also found, although
with a smaller amplitude than that seen in the continuum.  Lira et al. also
analysed broadband photometric images ({\em B} and {\em I} bands) and
detected variability of order 20 per cent between nights.
Optical spectroscopic data on NGC 4395 had also been obtained over a
number of years at Lick Observatory (Kraemer et al. 1999).  When these
data were investigated to determine whether they showed variability
similar to that seen by Lira et al., they found no evidence of the
factor of 2 variations but more modest variability could not be ruled
out (Moran et al. 1999). 

A well-studied correlation is seen to exist between the amplitude of
the variability in X-rays and the intrinsic luminosity of the nucleus
for local AGN (Manners et al. 2002; Almaini et al. 2000; Turner et al. 1999; Nandra et al.
1997), showing that higher luminosity AGN exhibit lower amplitude
variability.  In contrast, many low-luminosity AGN (typical X-ray
luminosity $0.4-60 \times 10^{40}$ ergs s$^{-1}$) in nearby large
galaxies show very little variability in X-rays, which has been
interpreted as evidence for an inefficient accretion process such as
an advection-dominated accretion flow (Ptak et al. 1998).  NGC 4395 has
been well studied at X-ray wavelengths and fast, large amplitude
variability has been observed in this object. Lira et al. (1999)
found variability of about a factor of 2 over a time-scale of 15 days
from \emph{ROSAT} PSPC (0.1-2.4 keV) data.  Such variability is not
particularly fast for Seyfert 1 AGN.  However, subsequent variability
studies with better sensitivity and time resolution in the harder 2-10 keV band using \emph{ASCA} found X-ray flux
changes on time-scales of about 100 seconds to 12 hours, with the
shortest doubling time-scale found to be of order 100s (Iwasawa et al. 2000).  Follow-up \emph{ASCA} observations (Shih et al. 2003) confirmed the earlier findings and estimated the power density
spectrum in the 1.2-10 keV and 2-10 keV bands, with the possible
detection of a break in the best-fitting power-law.  These findings
suggest that, while other dwarf Seyferts may have normal-size central
engines but radiatively inefficient accretion processes, NGC 4395 is
accreting at a normal fraction of the Eddington limit but genuinely
contains a small central engine when compared to other classical
Seyfert 1s.

To test the hypothesis that the Seyfert nucleus in NGC 4395 contains a
small central black hole, we have undertaken optical spectroscopic observations of NGC 4395 with the aim of monitoring the nuclear region every $\sim$ 30
minutes for several consecutive nights.  The motives for this study were several: firstly, it was important to check
whether the short time-scale variability observed in the broadband
optical observations of Lira et al. was characteristic of NGC 4395 or a
freak event, and secondly we wished to determine whether the
variability seen in NGC 4395 had different characteristics to
classical Seyfert 1s, in both amplitude and time-scale.  Although the
absolute flux calibration of spectroscopic observations is often less
accurate than for broadband photometry, a
careful observing procedure can reduce the uncertainties and allow
quantitative statements to be made.  The spectroscopic observations had two
advantages: if rapid continuum variability was observed, it could be determined
(1) whether there was a corresponding colour change and (2) whether any
simultaneous variability was seen in the broad lines.  If broad
line variability was seen, it was hoped that the time sampling of the
observations would allow any lag between the continuum and line
variations to be estimated, thus determining whether the size of the
broad-line region
scales with luminosity (Peterson et al. 2000). In this paper the detection of optical
continuum variability on time-scales $\sim$ 8 hours is reported.  In Sections 2
and 3 the observations and reduction of the optical spectra are
presented.   Section 4 describes the analysis procedure and checks
that were carried out to determine whether the observed variability
was due to systematic or atmospheric effects, and discusses the observed
variability of NGC 4395.  The results are discussed with respect to
other well-studied AGN and the implications for models of nuclear
regions in active galaxies in Section 5.  The subsequent conclusions
are summarised in Section 6.

\section{Observations}

Spectra of the nuclear region of NGC 4395 were obtained on 1998 March
17-20 using the 2.5-m Isaac Newton Telescope at the Roque de los
Muchachos Observatory.  A TEK 1024 $\times$ 1024 CCD was used together
with a R300V grating.  The CCD was windowed to give a scale of
0.7 arcsec/pixel  in the spatial direction and the
central wavelength of the observations was 5135\AA, giving a
dispersion of 3.32\AA/pixel over a wavelength range of $\sim 3500-6900$\AA.  

In order to detect spectral variations of the order of 20 per cent, a
meticulous observing strategy was required, since the standard
procedure rarely results in better than 30 per cent accuracy in the absolute
flux calibration.  Our aim in this study was for an accuracy of $\sim$
5 per cent, and to realise this the following procedure was implemented.

NGC 4395 was visible at an elevation of $> 30^{\rm{o}}$ for the
entirety of the four nights.  Conditions were cloudless for 3 of the
4 nights: the data from night 4 was affected by the presence of
high cirrus for part of the night and was scrapped.  Slit width was
alternated between 2 and 8 arcseconds, giving separate
interleaved sets of observations of NGC 4395, with a sampling interval of
approximately 30 minutes in both cases.  In all observations the slit
was positioned at the parallactic angle to prevent light losses due
to differential refraction through the atmosphere.  The use of two
slit widths allowed a number of systematic effects to be corrected for
or eliminated entirely.  The 2 arcsec slit was used
to reduce the contribution of the host galaxy starlight in each
spectrum, while the 8 arcsec slit was used to minimize
aperture effects due to variable seeing conditions, poor centring of
the object in the slit and the possibility that the narrow-line region
was resolved.  This last point is unlikely: analysis of
\emph{HST} images by both Filippenko et al. (1993) and Lira et al. (1999) found that the nucleus of NGC 4395 had an intrinsic
FWHM of less than 0.05 arcsec.  Seeing conditions were never better
than 1 arcsec during these observations, and so we can be confident
that resolution of the NLR is not a serious consideration in this
case.  Each NGC 4395 observation was bracketed by observations of
photometric standard stars (each observed using both slits at all
times).  These allowed any variations in both the seeing conditions
and atmospheric extinction to be quantified and if necessary corrected
for.  Bias frames, tungsten flats and twilight sky flats were taken at the
beginning and end of each night.  A log of the NGC 4395 observations is given in Table \ref{obs_journal}.


	\begin{table*}
	\caption{Journal of NGC 4395 observations}
	\label{obs_journal}
	\begin{center}
	\begin{tabular}{lcccc@{\hspace{23mm}}lcccc}
	\hline
	\hline
JD   &  Slit Width & Exp. Time  & PA &  Airmass & JD	& Slit Width & Exp. Time	& PA &
Airmass \\
{\footnotesize +2450800} & &{\footnotesize seconds} & & & {\footnotesize
+2450800} & &{\footnotesize seconds} & & \\
\hline
\noalign{\medskip}
\multicolumn{5}{c}{Night 1} & \multicolumn{5}{c}{Night 2, Continued} \\
\noalign{\smallskip} 
90.3959  & 2	&   500	&  104.5  &  1.791  & 91.6016	 &  2	&  1000	&  121.5  &  1.012  \\
90.4022 &  2	&  500	&  103.6  &  1.711  & 91.6116	&  8	&  200	&  111.1  &  1.021  \\
90.4090 &  8    &  100	&  102.6  &  1.633  & 91.6256	 &  2	&  1000	&  103.4  &  1.039  \\
90.4233  &  2	&  1000	&  100.4  &  1.496  & 91.6337	&  8	&  200	&  100.0  &  1.051  \\
90.4312 &  8	&  200	&  99.2	  &  1.431  & 91.6474	 &  2	&  1000	&  95.8	  &  1.079  \\
90.4453  &  2	&  1000	&  97.0	  &  1.335  & 91.6552	&  8	&  200	&  93.7	  &  1.098  \\
90.4544 &  8	&  200	&  95.6	  &  1.283  & 91.6684	 &  2	&  1000	&  90.8	  &  1.137  \\
90.4681  &  2	&  1000	&  93.2	  &  1.216  & 91.6762	&  8	&  200	&  89.2	  &  1.163  \\
90.4761 &  8	&  200	&  91.8	  &  1.183  & 91.6891	 &  2	&  1000	&  86.8	  &  1.215  \\
90.4896  &  2	&  1000	&  89.2	  &  1.135  & 91.6969	&  8	&  200	&  85.5	  &  1.250  \\
90.4995 &  8	&  200	&  87.0	  &  1.106  & 91.7025	 &  2	&  1000	&  83.4	  &  1.319  \\
90.5125  &  2	&  1000	&  83.6	  &  1.074  & 91.7175	&  8	&  200	&  82.2	  &  1.368  \\
90.5204 &  8	&  200	&  81.3	  &  1.058  & 91.7293	 &  2	&  1000	&  80.4	  &  1.454  \\
90.5336  &  2	&  1000	&  75.9	  &  1.036  & 91.7370	&  8	&  200	&  79.2	  &  1.519  \\
90.5429 &  8	&  200	&  71.0	  &  1.024  & 91.7503	 &  2	&  1000	&  77.2	  &  1.652  \\
90.5562  &  2	&  1000	&  58.3	  &  1.012  & 91.7581	&  8	&  200	&  77.2	  &  1.744  \\
90.5642 &  8	&  200	&  47.3	  &  1.007  & 91.7676	 &  2	&  1000	&  74.6	  &  1.875  \\
90.5763  &  2	&  1000	&  12.8	  &  1.004  & 91.7725	&  8	&  200	&  73.9	  &  1.952  \\
90.5842 &  8	&  200	&  165.1  &  1.004  & & & &  \\
90.5972  &  2	&  1000	&  132.5  &  1.008  & \multicolumn{5}{c}{Night 3}\\
90.6053 &  8	&  200	&  119.3  &  1.013  & & & &  \\
90.6765  & 2    &  500	&  89.7	  &  1.154  & 92.4082	 &  2	&  1000	&  101.9  &  1.586  \\
90.6828  & 2	&  500	&  88.5	  &  1.177  & 92.4262	&  8	&  200	&  99.1	  &  1.428  \\
90.6880 &  8	&  200	&  87.5	  &  1.198  & 92.4438	 &  2	&  1000	&  96.4	  &  1.312  \\
90.7000  &  2	&  1000	&  85.4	  &  1.253  & 92.4516	&  8	&  200	&  95.1	  &  1.269  \\
90.7080 &  8	&  200	&  84.1	  &  1.294  & 92.4644	 &  2	&  1000	&  92.9	  &  1.209  \\
90.7200  &  2	&  1000	&  82.2	  &  1.368  & 92.4724	&  8	&  200	&  91.5	  &  1.176  \\
90.7279 &  8	&  200	&  81.0	  &  1.422  & 92.4856	 &  2	&  1000	&  88.8	  &  1.131  \\
90.7398  &  2	&  1000	&  79.2	  &  1.520  & 92.4934	&  8	&  200	&  87.2	  &  1.107  \\
90.7475 &  8	&  200	&  78.0	  &  1.592  & 92.5086	 &  2	&  1000	&  83.2	  &  1.071  \\
90.7565	 & 2	&  500	&  76.7	  &  1.691  & 92.5164	&  8	&  200	&  80.8	  &  1.055  \\
90.7628	 & 2	&  500	&  75.8	  &  1.769  & 92.5301	 &  2	&  1000	&  74.9	  &  1.033  \\
90.7677	&  8	&  150	&  75.0	  &  1.835  & 92.5381	&  8	&  200	&  70.6	  &  1.023  \\
90.7701	 & 2	&  100	&  74.7	  &  1.870  & 92.5513	 &  2	&  1000	&  57.6	  &  1.012  \\
 & & &  & & 92.5592	&  8	&  200	&  46.4	  &  1.007  \\
\multicolumn{5}{c}{Night 2} & 92.5735	 &  2	&  1000	&  4.2	  &  1.004  \\
 & & &  & &  92.5813	&  8	&  200	&  156.5  &  1.004  \\
91.3987	 &  2	&  1000	&  103.7  &  1.721  & 92.5939	 &  2	&  1000	&  128.8  &  1.009  \\
91.4065	&  8	&  200	&  102.5  &  1.630  & 92.6016	&  8	&  200	&  117.5  &  1.014  \\
91.4235	 &  2	&  1000	&  100.0  &  1.471  & 92.6145	 &  2	&  1000	&  107.9  &  1.027  \\
91.4313	&  8	&  200	&  98.8	  &  1.410  & 92.6222	&  8	&  200	&  103.5  &  1.037  \\
91.4455	 &  2	&  1000	&  96.6	  &  1.318  & 92.6363	 &  2	&  1000	&  98.3	  &  1.061  \\
91.4534	&  8	&  200	&  95.3	  &  1.274  & 92.6441	&  8	&  200	&  95.9	  &  1.078  \\
91.4667	 &  2	&  1000	&  93.0	  &  1.211  & 92.6573	 &  2	&  1000	&  92.6	  &  1.111  \\
91.4746	&  8	&  200	&  91.6	  &  1.178  & 92.6650	&  8	&  200	&  90.9	  &  1.134  \\
91.4918	 &  2	&  1000	&  88.1	  &  1.120  & 92.6792	 &  2	&  1000	&  88.1	  &  1.185  \\
91.4998	&  8	&  200	&  86.3	  &  1.098  & 92.6870	&  8	&  200	&  86.7	  &  1.217  \\
91.5203	 &  2	&  1000	&  80.3	  &  1.053  & 92.7002	 &  2	&  1000	&  84.5	  &  1.282  \\
91.5284	&  8	&  200	&  77.2	  &  1.039  & 92.7079	&  8	&  200	&  83.2	  &  1.325  \\
91.5421	 &  2	&  1000	&  69.3	  &  1.022  & 92.7245	 &  2	&  1000	&  80.7	  &  1.439  \\
91.5499	&  8	&  200	&  63.3	  &  1.015  & 92.7324	&  8	&  200	&  79.5	  &  1.502  \\
91.5623	 &  2	&  1000	&  43.8	  &  1.007  & 92.7452	 &  2	&  1000	&  77.6	  &  1.626  \\
91.5701	&  8	&  200	&  25.9	  &  1.004  & 92.7531	&  8	&  200	&  76.4	  &  1.716  \\
91.5822	 &  2	&  1000	&  164.4  &  1.004  & 92.7661	 &  2	&  1000	&  74.4	  &  1.895  \\
91.5900	&  8	&  200	&  140.0  &  1.006  & 92.7723	&  8	&  200	&  73.5	  &  1.994  \\
\hline
\hline
	\end{tabular}
	\end{center}
	\end{table*}

\section{Data reduction}
The data were reduced using {\sc iraf} software routines, and bias
correction was performed in the standard way.  Pixel-to-pixel
sensitivity variations were removed using tungsten lamp spectra, which
were observed to contain vignetting effects.  It was decided to
correct these effects using spectra of the twilight sky, as the
optical light path for these frames was the same as for the spectra of
NGC 4395 itself.  Therefore the tungsten spectrum was first smoothed
using a median filter designed to smooth the gain variations but
preserve the vignetting features.  The unsmoothed flat was then
divided through by the smoothed version to produce a flat field that
contained information about the pixel-to-pixel variations but no
information about the vignetting.  Each twilight sky and NGC
4395 spectrum was divided through by this frame.  The
vignetting effects were then removed from the NGC 4395 spectrum by
dividing through by the processed twilight sky frame that containing
the vignetting information. 

For wavelength calibration, a low-order polynomial function was fitted
to strong lines in a copper-argon comparison spectrum.  The precision
of the fit was improved by including intermediate lines once a rough
determination of the dispersion solution had been found.  Dispersion
solutions were found every 35 arcsec (50 pixels) along the slit and
6th-order Chebyshev polynomials were fitted to these solutions to correct
any geometrical distortions.  To remove the sky background, a low-order cubic spline
function was fitted along the spatial axis of each spectrum and then
subtracted.  One-dimensional spectra were extracted using a $\sim 5$ arcsec
aperture designed to minimize any contributions from the host galaxy,
including nearby {\sc hii} regions.  Lira et al. (1999) estimated
the contamination from host galaxy starlight to be at most 10 per cent from
measurements of the equivalent width of the ultraviolet Ca {\sc ii} line.

\begin{figure}
	\centerline{\psfig{figure=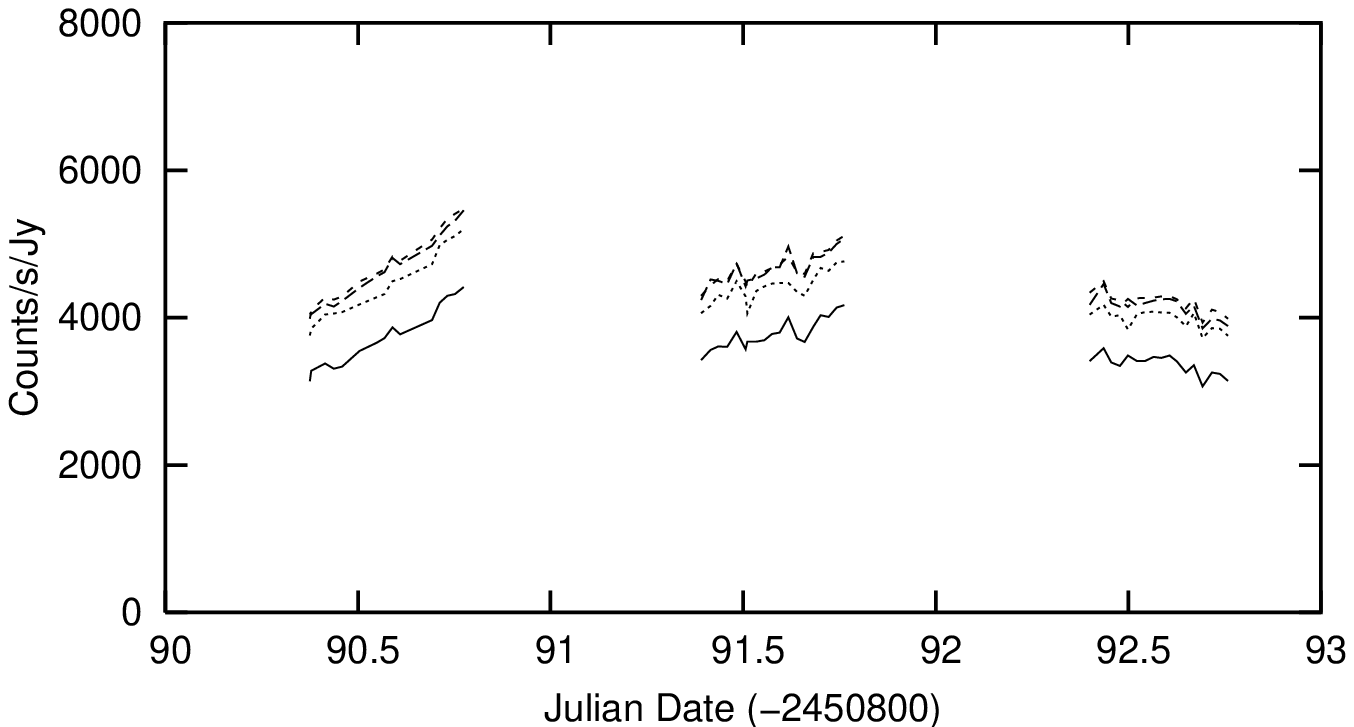,width=8.0cm}}
	\caption{The variation in counts sec$^{-1}$ Jy$^{-1}$ over
the three photometric nights for the standard stars
observed using the 8 arcsec slit.  To ensure that any variations were
grey, measurements were made at four wavelengths: 4000{\AA} (solid line),
4500{\AA} (short dashes), 5500{\AA} (long dashes) and 6200{\AA} (dotted).}
	\label{seeing_vars}
	\end{figure}

Flux calibration of the extracted NGC 4395 spectra was initially
performed in the standard way.  Spectrophotometric standard stars were
used to determine the sensitivity of the detector over the entire
wavelength range.  The standard stars used were
Feige 34, G191-B2B and HZ44 from Massey et al. (1988), together with
BD75+325 and HZ21 from Oke (1990).  The standard extinction curve for the Roque de los Muchachos
observatory was used to perform the extinction correction.  For each
NGC 4395 observation the sensitivity function was constructed using
the standard star observations closest in time and using the same slit
width.  Since a $\sim$ 5 per cent accuracy was desired, in order to ensure
that the extinction correction had been applied correctly the standard
extinction correction was applied to each standard star observation,
and then the counts sec$^{-1}$ Jy$^{-1}$ was measured at 5500{\AA}
(Johnson \emph{V}).  Significant variations were seen in the derived
calibration both within a night and on a night-to-night basis.  To correct
for the night-to-night variations, a grey shift correction was applied
to all spectra (including both spectrophotometric standards and NGC
4395 observations) based on measurements made on-site by the Carlsberg
Meridian Circle (CMC).  The grey shifts for the three nights were
0.30, 0.20 and 0.13 magnitudes respectively.  To correct for
atmospheric variations within each night, the following procedure was
implemented.  First, the standard star observations taken with the 8
arcsec slit width were corrected to first order using the data from the CMC and the counts sec$^{-1}$
Jy$^{-1}$ for each observation were measured. This was done at several
different wavelengths to check that any intra-night variations
were grey, as shown in Figure \ref{seeing_vars}.  The mean of all
measurements (at one wavelength) was found and the magnitude shift
required to bring each observation to the mean value calculated.  The
8 arcsec slit data was used for this purpose because due to the large
slit width we could be reasonably confident that any variations seen
were solely due to variations in atmospheric transparency and
\emph{not} due to other effects such as seeing changes.  The magnitude
shifts required to apply the same second-order correction to the NGC
4395 data (both 2 and 8 arcsec slit widths) were then calculated by
interpolating between the 8 arcsec standard star observations, and
individual extinction curves were constructed for each observation.
The flux calibration was then repeated using a global sensitivity
function calculated for each night and the separate extinction
curves.  The procedure we have followed assumes that the transparency
trends we see in the standard stars are also followed by NGC 4395.
The smoothness of the trends suggests this is correct.  Any remaining
small time-scale transparency fluctuations could produce errors in our
derived light curves of NGC 4395.  We estimated this by fitting
straight lines to the data for each night in Figure \ref{seeing_vars},
and then calculating the residual variations.  This showed that any
such short time-scale transparency variations are less than 5 per cent (2$\sigma$).  

\begin{figure*}
	\begin{center}
	\centerline{\psfig{figure=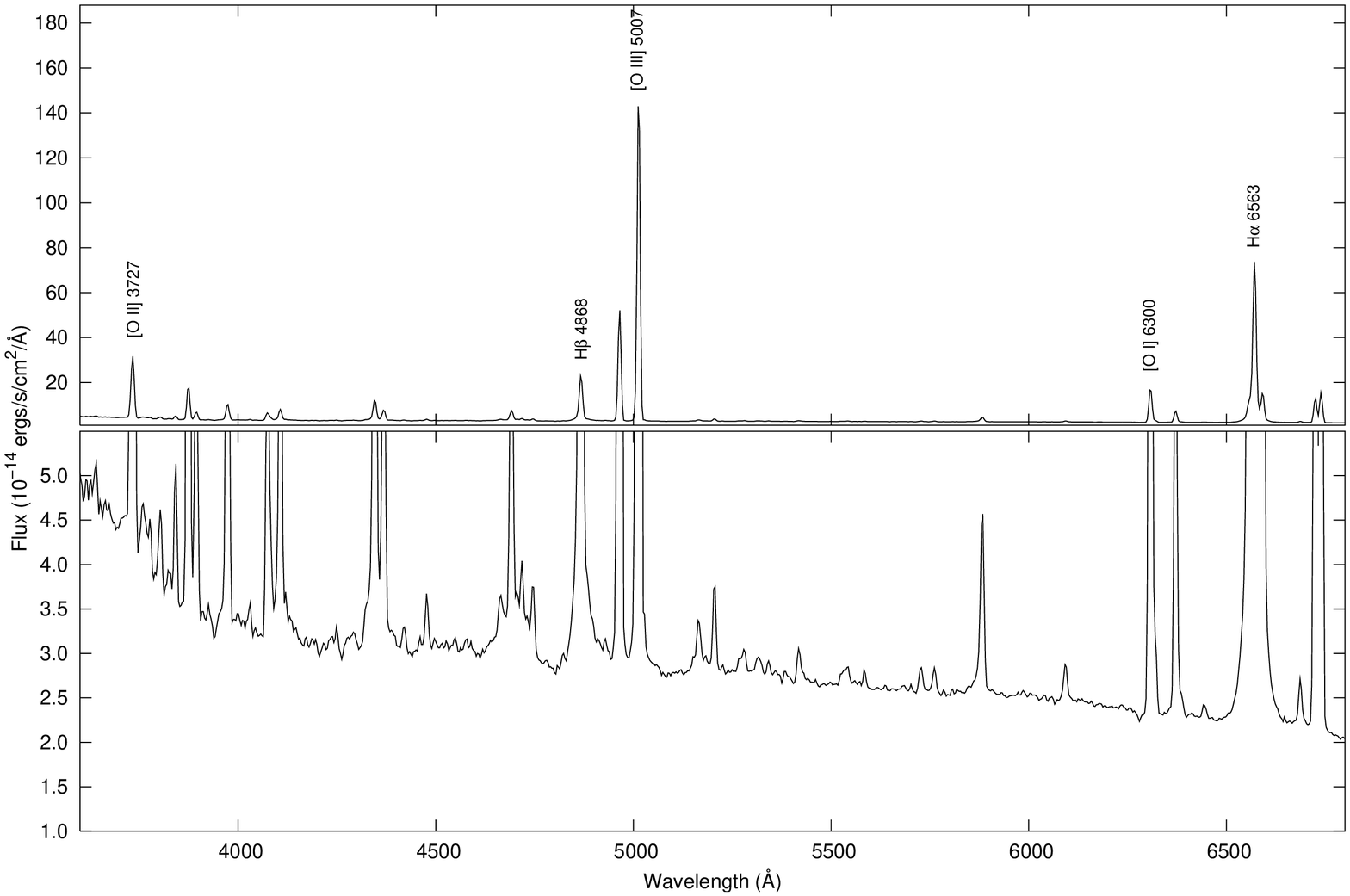,width=17cm}}
	\caption{Averaged spectrum for all observations taken using
the 2 arcsec slit.  The bottom panel has an expanded flux scale
showing the low S/N lines.}
	\label{summed_spec}
	\end{center}
	\end{figure*}

The combined spectrum is shown in Figure \ref{summed_spec}.  This has
extremely high signal-to-noise and many weak lines can be clearly
seen.  Most of the structure visible in the lower panel of Figure
\ref{summed_spec} is real rather than noise.  This spectrum will be discussed in a
later paper.


\section{Analysis and Results}

\subsection{Procedure}

We examined the continuum variability of NGC 4395 in two ways.  First,
using the absolute flux calibration described above, including the 2nd-order
transparency corrections; and second, by measuring the equivalent
widths of narrow lines, which we expect to be particularly reliable.  If the size of the
narrow-line region is of order $\sim$10 pc (Filippenko et al. 1993), this gives a light-crossing time in excess of 30 years, and so
we would not expect to see any significant variability in the narrow lines on time-scales
of hours or days.  If this is the case, then measurements of the
narrow line equivalent widths give a direct measurement of any continuum variability.  A further
advantage of this technique is that in the absence of systematic
effects the EQW is independent of the absolute flux calibration, and
so the principle sources of uncertainty will be the accuracies to
which we can measure the relative line and continuum fluxes.  Possible
sources of systematic errors and aperture effects in the NGC 4395 data
have been investigated and are discussed in the next section.

The equivalent widths were obtained by fitting low-order polynomial
functions to the continuum region under the emission lines, and then
fitting a Gaussian profile to the emission lines themselves.  The
emission lines selected were [O {\sc ii}] 3727{\AA}, [O {\sc iii}]
5007{\AA} and [O {\sc i}] 6300{\AA}.  These lines were selected
because they were the strongest unblended lines in the blue, green and
red regions of
the spectrum.  Several wavelength regions were required to allow
differences in variability with colour to be quantified.  To
measure the continuum flux, co-ordinate ranges close to the relevant
emission line were selected that would result in the least
contamination of the fit by low signal-to-noise emission lines, and a
polynomial of order zero or one was fitted to these co-ordinate
ranges.  The actual fitting order used was determined by eye.  A
Gaussian function was then fitted to the emission line to obtain the narrow
line flux and hence the equivalent width. 

The largest uncertainty in the equivalent widths arises
from the continuum flux measurement, particularly for the [O {\sc ii}]
3727{\AA} line which has the poorest signal-to-noise ratio in both
the 2 arcsec and 8 arcsec slit data sets.  Two methods were used to quantify
this error.  First, the continuum fit was manually constrained to lie some
distance away from the actual best-fitting value, and the effect upon
the equivalent width measurement observed.  Secondly, the co-ordinate
ranges used for the fitting were changed by small amounts, thus
varying the contribution from any low signal-to-noise emission lines,
and again the effect upon the equivalent widths determined.  The
actual uncertainty in the measurement is likely to lie somewhere
between these two values, with the fractional error being slightly
larger for the [O {\sc ii}] 3727{\AA} line than either of the other
two lines.

\begin{figure*}
	\centerline{\psfig{figure=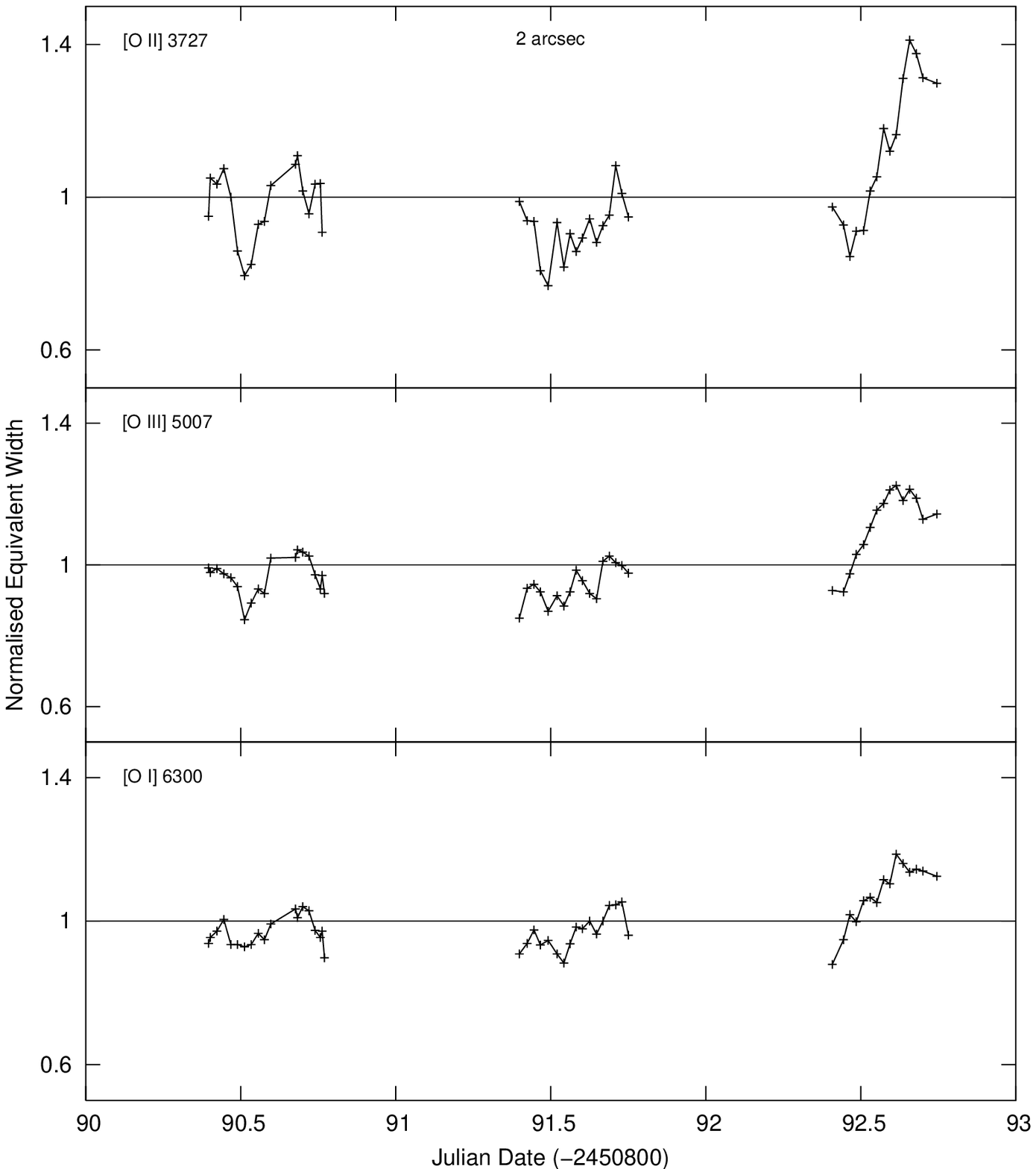,width=13cm}}
	\caption[]{Equivalent widths (scaled using mean equivalent
width) versus JD for the [{\sc o ii}] 3727{\AA} (top panel), [{\sc o iii}]
5007{\AA} (middle) and [{\sc o i}] 6300{\AA} (bottom) lines measured using from the 2
arcsec slit spectra.}
	\label{lc_2_all_scaled}
	\end{figure*}

\begin{figure*}
	\centerline{\psfig{figure=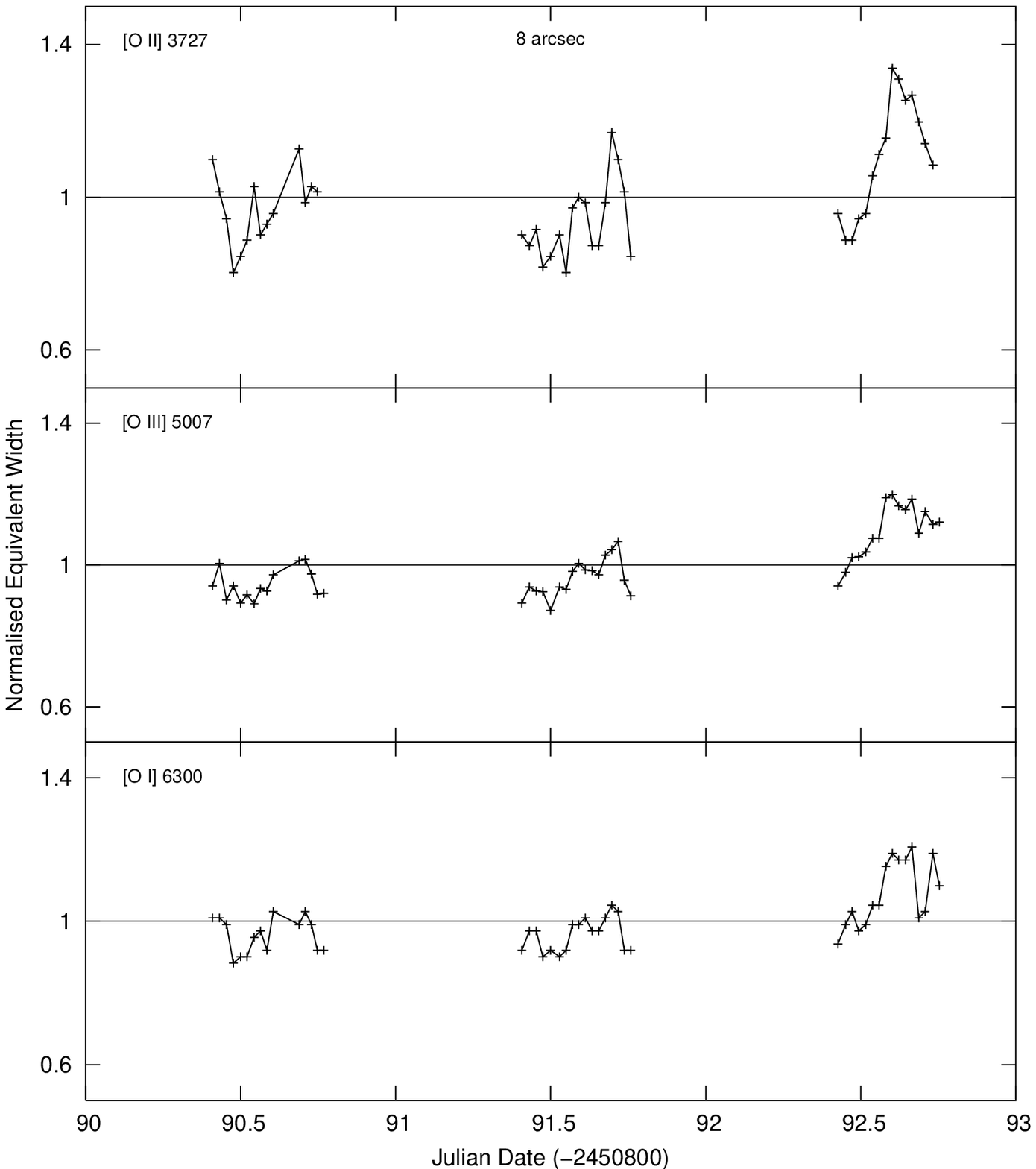,width=13cm}}
	\caption[]{Equivalent widths (scaled using mean equivalent
width) versus JD for the [{\sc o ii}] 3727{\AA} (top panel), [{\sc o iii}]
5007{\AA} (middle) and [{\sc o i}] 6300{\AA} (bottom) lines measured
using from the 8 arcsec slit spectra.}
	\label{lc_8_all_scaled}
	\end{figure*}

Table
\ref{flux_meas_2} gives 51 measurements for the equivalent widths and
the continuum and line fluxes for the [O {\sc iii}] 5007{\AA} and
[O {\sc i}] 6300{\AA} emission lines, as measured from the 2 arcsec
slit data.  There are only 50 measurements
for the [O {\sc ii}] 3727{\AA} line, since the signal-to-noise ratio
of the last observation on night 1 was too poor in the blue end to
allow the measurements to be made.  Table \ref{flux_meas_8} contains
the measurements made using data from the 8 arcsec slit measurements.
There are 48 measurements for both the [O {\sc iii}] 5007{\AA} and
[O {\sc i}] 6300{\AA} lines, but similarly only 47 for the [O {\sc ii}] 3727{\AA} line.  The final light curves produced from these data
are shown in Figures \ref{lc_2_all_scaled} (2 arcsec slit) and
\ref{lc_8_all_scaled} (8 arcsec slit).  To allow the measurements from
the different emission lines to be compared directly, all light curves
have been scaled using the mean equivalent width for each line.

\begin{figure*}
	\centerline{\psfig{figure=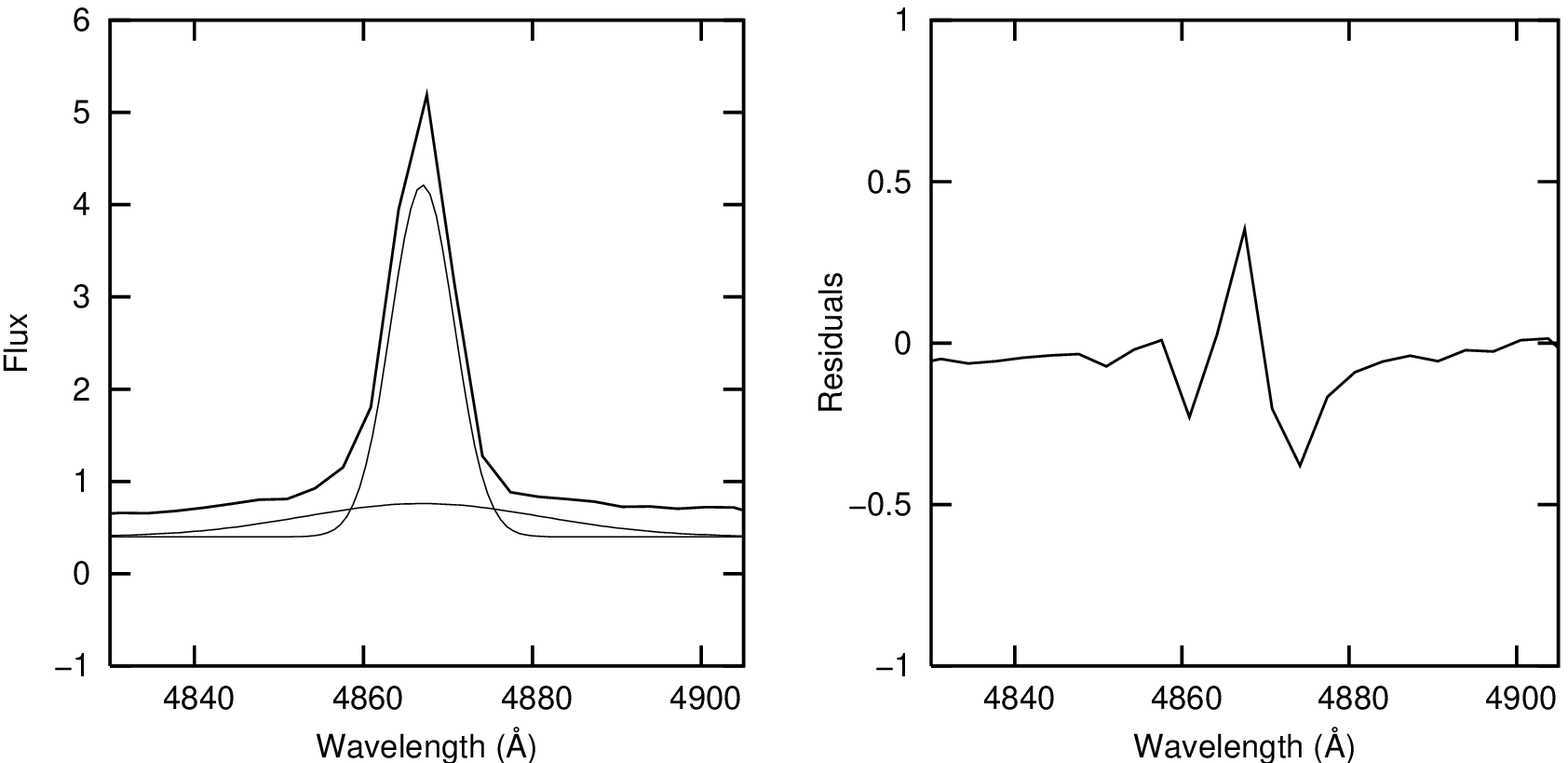,width=12.5cm}}
	\caption[]{Example Gaussian fit to H$\beta$.  The width of the
narrow component had a fixed instrumental width measured from other
lines, while the width of the broad component varied as a free
parameter.  All fluxes are in units of 10$^{-15}$ erg cm$^{-2}$
s$^{-1}$ \AA$^{-1}$.}
	\label{2_hbeta_exfit}
	\end{figure*}

Measurements of the H$\beta$ line flux were
also carried out, and a similar procedure to that described in
Lira et al. (1999) was followed.  First, the zero and slope of the
continuum region under the H$\beta$ line were determined by eye, and
then these parameters were fixed during the fitting process.  The line
itself was fitted using two Gaussians.  The narrow component of the line
was fitted using a Gaussian of fixed instrumental width determined from
the nearby narrow lines, with the amplitude and centre as free
parameters.  The broad component was fitted using a Gaussian in which the
amplitude, width and centre were all free parameters.  This procedure
was only possible for the 2 arcsec slit data, since the poorer
signal-to-noise of the 8 arcsec slit data due to starlight
contamination from the host galaxy meant that it was difficult
to distinguish the weak broad wings on the H$\beta$ line from the
continuum.  The measured narrow- and broad-component fluxes are given
in columns (11) and (12) of Table \ref{flux_meas_2},and an example fit
and residuals is shown in Figure \ref{2_hbeta_exfit}.  The measured
FWHM for the narrow and broad components were $\sim$250 km s$^{-1}$
(fixed using the instrumental resolution) and
$\sim$1800 km s$^{-1}$ respectively.  Previous studies give values of
$\leq$50 km s$^{-1}$ for the narrow component using high-resolution
spectroscopy (Filippenko \& Sargent 1989), and $\sim$1500 km s$^{-1}$
for the broad component (Kraemer et al. 1999).


	\begin{table*}
	\caption{Equivalent width, line flux and continuum flux
measurements for the 2 arcsec slit data.}
	\begin{center}
	\begin{tabular}{l@{\hspace{11mm}}ccc@{\hspace{11mm}}ccc@{\hspace{11mm}}ccc@{\hspace{11mm}}cc}

\hline
\hline

JD	& \multicolumn{3}{c}{[{\sc oii}] $\lambda$3727\AA} &
\multicolumn{3}{c}{[{\sc oiii}] $\lambda$5007\AA} &
\multicolumn{3}{c}{[{\sc oi}] $\lambda$6300\AA}&
\multicolumn{2}{c}{H$\beta$}\\
\footnotesize{+2450800} & EQW & Line$^1$ & Cont$^2$ & EQW & Line & Cont& EQW& Line & Cont & Narrow & Broad\\
\hline
90.3959311	& 58.6 & 4.558 & 7.775 & 469 & 22.80 & 4.861 & 57.7 & 2.325 & 4.027 & 1.827 & 0.935 \\  
90.4022447	& 64.8 & 4.946 & 7.637 & 463 & 23.95 & 5.168 & 58.7 & 2.487 & 4.235 & 2.005 & 0.910 \\
90.4233443	& 63.8 & 5.726 & 8.974 & 468 & 28.42 & 6.075 & 59.8 & 2.914 & 4.869 & 2.400 & 1.002 \\
90.4453813	& 66.3 & 5.463 & 8.243 & 461 & 25.60 & 5.551 & 61.8 & 2.538 & 4.108 & 2.085 & 1.071 \\
90.4681417	& 61.7 & 5.674 & 9.195 & 456 & 27.15 & 5.957 & 57.5 & 2.823 & 4.912 & 2.189 & 0.949 \\
90.4896174	& 53.0 & 5.645 & 10.65 & 444 & 29.69 & 6.694 & 57.5 & 3.156 & 5.491 & 2.424 & 1.058 \\
90.5125167	& 49.0 & 5.326 & 10.86 & 400 & 26.81 & 6.703 & 57.1 & 3.111 & 5.451 & 2.485 & 1.054 \\
90.5336336	& 50.8 & 4.950 & 9.738 & 422 & 26.82 & 6.359 & 57.5 & 3.007 & 5.232 & 2.440 & 0.980 \\
90.5562956	& 57.3 & 5.374 & 9.374 & 441 & 28.37 & 6.426 & 59.4 & 3.189 & 5.366 & 2.505 & 1.066 \\
90.5763477	& 57.8 & 5.267 & 9.105 & 435 & 26.73 & 6.147 & 58.3 & 2.979 & 5.110 & 2.401 & 0.980 \\
90.5972456	& 63.6 & 4.924 & 7.748 & 482 & 26.67 & 5.533 & 61.0 & 3.116 & 5.111 & 2.276 & 0.965 \\
90.6765387	& 67.0 & 6.263 & 9.353 & 483 & 27.74 & 5.748 & 63.6 & 2.864 & 4.500 & 2.363 & 0.951 \\
90.6828755	& 68.4 & 5.857 & 8.559 & 493 & 27.23 & 5.527 & 62.1 & 2.745 & 4.421 & 2.181 & 1.103 \\
90.7000862	& 62.7 & 5.737 & 9.152 & 490 & 28.36 & 5.788 & 64.0 & 2.932 & 4.584 & 2.426 & 1.008 \\
90.7200977	& 59.0 & 5.479 & 9.280 & 485 & 29.07 & 5.994 & 63.3 & 2.927 & 4.625 & 2.471 & 1.110 \\
90.7398720	& 63.8 & 5.763 & 9.030 & 460 & 27.22 & 5.915 & 59.9 & 2.828 & 4.725 & 2.225 & 1.093 \\
90.7565387	& 63.9 & 5.343 & 8.362 & 441 & 23.49 & 5.329 & 58.7 & 2.462 & 4.195 & 1.889 & 0.991 \\
90.7628350	& 56.0 & 5.432 & 9.693 & 459 & 24.89 & 5.419 & 59.8 & 2.606 & 4.360 & 1.950 & 1.032 \\
90.7701035	&      &       &       & 435 & 23.33 & 5.386 & 55.2 & 2.416 & 4.373 &  &  \\
\\		
91.3987031	& 61.0 & 3.997 & 6.556 & 402 & 19.84 & 4.937 & 55.9 & 2.429 & 4.343 & 1.815 & 0.935 \\
91.4235642	& 57.9 & 5.038 & 8.708 & 442 & 25.25 & 5.716 & 57.7 & 2.699 & 4.674 & 2.122 & 1.018 \\
91.4455433	& 57.8 & 5.008 & 8.670 & 447 & 24.99 & 5.592 & 60.0 & 2.581 & 4.304 & 2.230 & 1.001 \\
91.4667008	& 49.8 & 4.504 & 9.042 & 437 & 25.82 & 5.906 & 57.4 & 2.628 & 4.593 & 2.246 & 1.058 \\
91.4918165	& 47.4 & 4.521 & 9.543 & 411 & 24.81 & 6.042 & 58.2 & 2.705 & 4.646 & 2.294 & 1.041 \\
91.5203003	& 57.6 & 5.430 & 9.421 & 432 & 26.20 & 6.066 & 55.9 & 2.704 & 4.840 & 2.259 & 1.083 \\
91.5421116	& 50.4 & 4.608 & 9.140 & 418 & 24.06 & 5.758 & 54.3 & 2.396 & 4.415 & 2.221 & 1.083 \\
91.5623142	& 55.8 & 5.045 & 9.048 & 437 & 24.82 & 5.679 & 57.6 & 2.450 & 4.256 & 2.246 & 1.038 \\
91.5822621	& 52.9 & 4.834 & 9.141 & 466 & 27.06 & 5.810 & 60.5 & 2.711 & 4.481 & 2.433 & 0.946 \\
91.6016718	& 55.1 & 4.786 & 8.692 & 452 & 25.34 & 5.603 & 60.2 & 2.865 & 4.757 & 2.227 & 1.001 \\
91.6256822	& 58.2 & 4.885 & 8.390 & 435 & 24.74 & 5.692 & 61.5 & 2.898 & 4.715 & 2.196 & 1.045 \\
91.6474646	& 54.4 & 4.824 & 8.861 & 428 & 23.29 & 5.443 & 59.3 & 2.619 & 4.417 & 2.215 & 1.030 \\
91.6684774	& 57.1 & 5.483 & 9.601 & 478 & 29.28 & 6.121 & 61.5 & 3.029 & 4.923 & 2.564 & 1.080 \\
91.6891660	& 58.8 & 5.159 & 8.769 & 485 & 26.28 & 5.417 & 64.2 & 2.802 & 4.362 & 2.239 & 1.024 \\
91.7096753	& 66.8 & 4.850 & 7.263 & 476 & 23.58 & 4.957 & 64.3 & 2.568 & 3.995 & 2.011 & 0.983 \\
91.7293165	& 62.3 & 5.259 & 8.436 & 472 & 27.07 & 5.736 & 64.8 & 3.065 & 4.733 & 2.205 & 1.138 \\
91.7503119	& 58.5 & 4.916 & 8.399 & 462 & 24.80 & 5.368 & 59.1 & 2.599 & 4.399 & 2.097 & 0.957 \\
	\\
92.4082054	& 60.1 & 5.293 & 8.799 & 439 & 26.54 & 6.049 & 54.1 & 2.827 & 5.223 & 2.232 & 1.075 \\
92.4438362	& 57.2 & 5.267 & 9.203 & 437 & 26.89 & 6.152 & 58.3 & 2.944 & 5.049 & 2.513 & 1.103 \\
92.4644959	& 52.1 & 5.063 & 9.714 & 461 & 28.51 & 6.181 & 62.6 & 2.999 & 4.792 & 2.633 & 1.216 \\
92.4856764	& 56.2 & 5.014 & 8.914 & 487 & 27.28 & 5.598 & 61.4 & 2.848 & 4.636 & 2.513 & 1.072 \\
92.5086510	& 56.3 & 4.930 & 8.763 & 500 & 27.79 & 5.563 & 65.0 & 2.946 & 4.531 & 2.556 & 1.086 \\
92.5301961	& 62.7 & 5.185 & 8.275 & 523 & 28.85 & 5.514 & 65.6 & 2.905 & 4.431 & 2.549 & 1.045 \\
92.5513535	& 65.0 & 5.152 & 7.929 & 546 & 29.13 & 5.334 & 64.7 & 2.827 & 4.372 & 2.690 & 1.081 \\
92.5735642	& 72.8 & 5.080 & 6.979 & 555 & 26.85 & 4.752 & 68.6 & 2.700 & 3.934 & 2.360 & 1.006 \\
92.5939288	& 69.1 & 5.268 & 7.627 & 573 & 29.02 & 5.065 & 67.9 & 2.860 & 4.214 & 2.476 & 1.041 \\
92.6145017	& 71.8 & 5.686 & 7.918 & 579 & 29.87 & 5.157 & 73.0 & 3.036 & 4.161 & 2.504 & 1.043 \\
92.6363651	& 80.9 & 6.584 & 8.140 & 559 & 31.49 & 5.635 & 71.4 & 3.232 & 4.524 & 2.682 & 1.092 \\
92.6573258	& 87.1 & 6.006 & 6.893 & 574 & 27.40 & 4.843 & 69.9 & 2.777 & 3.974 & 2.224 & 0.973 \\
92.6792123	& 84.9 & 7.328 & 8.634 & 562 & 32.68 & 5.814 & 70.4 & 3.390 & 4.817 & 2.713 & 1.045 \\
92.7002193	& 81.0 & 7.207 & 8.893 & 534 & 32.69 & 6.120 & 70.1 & 3.563 & 5.085 & 2.588 & 1.111 \\
92.7452077	& 80.1 & 5.399 & 6.742 & 541 & 26.84 & 4.960 & 69.2 & 2.794 & 4.038 & 2.130 & 0.983 \\
\hline
\hline
\multicolumn{12}{l}{$^1$Line fluxes in units of $10^{-14}$ ergs s$^{-1}$ cm$^{-2}$}\\
\multicolumn{12}{l}{$^2$Continuum fluxes in units of $10^{-16}$ ergs
s$^{-1}$ cm$^{-2}${\AA}$^{-1}$}
	\end{tabular}
	\end{center}
	\label{flux_meas_2}
	\end{table*}


	\begin{table*}
	\caption{Equivalent widths, continuum and line flux
measurements for 8 arcsec slit observations.}
	\begin{center}
	\begin{tabular}{l@{\hspace{10mm}}ccc@{\hspace{13mm}}ccc@{\hspace{13mm}}ccc}

\hline
\hline

JD	& \multicolumn{3}{c}{[{\sc oii}] $\lambda$3727\AA} &
\multicolumn{3}{c}{[{\sc oiii}] $\lambda$5007\AA} &
\multicolumn{3}{c}{[{\sc oi}] $\lambda$6300\AA} \\

\footnotesize{+2450800} & EQW & Line$^1$ & Cont$^2$ & EQW & Line & Cont & EQW
& Line & Cont \\
\hline
90.4090040	& 77.9 & 7.020 & 9.015 & 411 & 27.39 & 6.660 & 56.4 & 3.031 & 5.376 \\
90.4312551	& 72.0 & 6.254 & 8.685 & 439 & 27.60 & 6.291 & 56.3 & 3.011 & 5.351 \\
90.4544959	& 66.8 & 6.492 & 9.719 & 394 & 27.29 & 6.921 & 54.8 & 3.212 & 5.861 \\
90.4761452	& 56.8 & 5.800 & 10.21 & 411 & 29.15 & 7.098 & 48.6 & 3.144 & 6.471 \\
90.4995480	& 60.4 & 6.721 & 11.12 & 390 & 29.10 & 7.454 & 49.7 & 3.111 & 6.263 \\
90.5204855	& 65.7 & 6.812 & 10.37 & 400 & 29.74 & 7.433 & 50.5 & 3.200 & 6.339 \\
90.5429507	& 73.9 & 7.468 & 10.11 & 389 & 28.58 & 7.343 & 53.2 & 3.291 & 6.184 \\
90.5642586	& 64.1 & 6.410 & 10.00 & 408 & 28.22 & 6.924 & 54.4 & 3.147 & 5.782 \\
90.5842470	& 65.8 & 6.561 & 9.976 & 405 & 28.04 & 6.920 & 51.4 & 3.025 & 5.880 \\
90.6053176	& 67.6 & 6.286 & 9.292 & 425 & 29.37 & 6.907 & 56.8 & 3.348 & 5.894 \\
90.6880549	& 80.4 & 6.263 & 7.793 & 442 & 27.90 & 6.316 & 55.5 & 3.102 & 5.590 \\
90.7080896	& 70.2 & 6.074 & 8.650 & 444 & 27.59 & 6.208 & 56.5 & 3.126 & 5.529 \\
90.7279623	& 72.6 & 6.462 & 8.896 & 426 & 27.79 & 6.522 & 55.0 & 3.013 & 5.475 \\
90.7475341	& 71.9 & 6.593 & 9.166 & 401 & 28.49 & 7.107 & 51.4 & 3.123 & 6.076 \\
90.7677135	&  &       &       & 402 & 27.71 & 6.887 & 51.2 & 2.987 & 5.838 \\
	\\
91.4065966	& 64.4 & 6.564 & 10.19 & 390 & 27.12 & 6.948 & 51.4 & 2.968 & 5.778 \\
91.4313883	& 62.0 & 5.565 & 8.971 & 410 & 25.63 & 6.246 & 54.1 & 2.791 & 5.159 \\
91.4534484	& 65.1 & 6.215 & 9.551 & 405 & 27.55 & 6.802 & 53.7 & 2.966 & 5.520 \\
91.4746001	& 57.9 & 6.084 & 10.50 & 404 & 26.93 & 6.665 & 49.6 & 2.817 & 5.685 \\
91.4998489	& 60.0 & 6.286 & 10.48 & 381 & 27.22 & 7.146 & 51.3 & 2.929 & 5.705 \\
91.5284947	& 64.4 & 5.978 & 9.282 & 410 & 27.43 & 6.683 & 50.0 & 2.902 & 5.809 \\
91.5499415	& 57.5 & 5.632 & 9.789 & 407 & 27.06 & 6.652 & 51.0 & 2.904 & 5.692 \\
91.5701730	& 69.1 & 6.534 & 9.450 & 429 & 26.85 & 6.255 & 55.0 & 2.995 & 5.444 \\
91.5900283	& 70.6 & 6.198 & 8.779 & 439 & 26.88 & 6.118 & 54.6 & 2.945 & 5.392 \\
91.6116892	& 69.6 & 6.429 & 9.234 & 431 & 27.18 & 6.300 & 56.0 & 3.020 & 5.391 \\
91.6337262	& 61.5 & 5.822 & 9.463 & 430 & 26.88 & 6.254 & 53.5 & 2.846 & 5.319 \\
91.6552135	& 62.0 & 5.940 & 9.580 & 425 & 26.95 & 6.340 & 54.5 & 3.078 & 5.651 \\
91.6762725	& 70.4 & 5.612 & 7.977 & 449 & 25.98 & 5.780 & 56.1 & 2.827 & 5.041 \\
91.6969612	& 82.6 & 6.332 & 7.667 & 456 & 26.55 & 5.821 & 59.9 & 3.030 & 5.223 \\
91.7175572	& 78.2 & 6.221 & 7.959 & 466 & 26.04 & 5.583 & 56.6 & 2.827 & 4.997 \\
91.7370827	& 72.0 & 6.132 & 8.514 & 418 & 27.33 & 6.535 & 50.7 & 2.890 & 5.701 \\
91.7581764	& 59.9 & 6.266 & 10.46 & 399 & 28.19 & 7.061 & 50.5 & 3.007 & 5.949 \\
	\\
92.4262725	& 67.7 & 5.688 & 8.400 & 411 & 26.05 & 6.345 & 52.2 & 2.707 & 5.185 \\
92.4516776	& 63.5 & 6.100 & 9.610 & 428 & 27.80 & 6.494 & 54.9 & 2.833 & 5.160 \\
92.4724531	& 63.2 & 6.223 & 9.844 & 446 & 28.63 & 6.413 & 56.7 & 2.905 & 5.120 \\
92.4934542	& 66.9 & 6.119 & 9.142 & 447 & 29.16 & 6.521 & 54.2 & 2.873 & 5.304 \\
92.5164461	& 68.2 & 6.082 & 8.921 & 453 & 29.54 & 6.522 & 55.1 & 2.878 & 5.226 \\
92.5381012	& 74.5 & 6.409 & 8.602 & 470 & 29.49 & 6.270 & 58.0 & 3.007 & 5.181 \\
92.5592702	& 79.1 & 5.986 & 7.563 & 470 & 27.75 & 5.901 & 58.2 & 2.797 & 4.808 \\
92.5813998	& 82.3 & 6.382 & 7.751 & 520 & 27.68 & 5.318 & 65.5 & 2.824 & 4.311 \\
92.6016660	& 94.5 & 6.392 & 6.762 & 524 & 27.22 & 5.198 & 66.3 & 2.729 & 4.115 \\
92.6222563	& 93.0 & 6.451 & 6.933 & 510 & 28.06 & 5.499 & 65.1 & 2.842 & 4.365 \\
92.6441545	& 89.3 & 6.687 & 7.485 & 505 & 27.66 & 5.479 & 64.6 & 2.933 & 4.537 \\
92.6650746	& 89.8 & 5.994 & 6.672 & 518 & 25.71 & 4.959 & 66.6 & 2.701 & 4.058 \\
92.6870017	& 85.8 & 6.848 & 7.983 & 476 & 28.52 & 5.987 & 56.2 & 2.721 & 4.841 \\
92.7079970	& 80.8 & 5.447 & 6.742 & 503 & 25.41 & 5.054 & 57.5 & 2.464 & 4.287 \\
92.7324183	& 76.7 & 5.171 & 6.740 & 487 & 24.54 & 5.044 & 66.3 & 2.643 & 3.987 \\
92.7531591	&  &       &       & 490 & 26.29 & 5.364 & 61.2 & 2.710 & 4.425 \\
\hline
\hline
\multicolumn{10}{l}{$^1$Line fluxes in units of $10^{-14}$ ergs s$^{-1}$ cm$^{-2}$}\\
\multicolumn{10}{l}{$^2$Continuum fluxes in units of $10^{-16}$ ergs
s$^{-1}$ cm$^{-2}${\AA}$^{-1}$}
	\end{tabular}
	\end{center}
	\label{flux_meas_8}
	\end{table*}


\subsection{Diagnostic Checks}

Figures \ref{lc_2_all_scaled} and \ref{lc_8_all_scaled} show short
time-scale variations of the nucleus of NGC 4395 throughout all 3 nights.  In order to determine
whether these variations were real intrinsic variations of the source or due to aperture or systematic
effects a number of diagnostic checks were implemented.

Although NGC 4395 is a very low surface-brightness galaxy, it
nonetheless shows a number of isolated knots of star formation and H {\sc ii} regions.  Since the slit was positioned
always at the parallactic angle, this resulted in the position angle
of the slit relative to the galaxy rotating over the course of the
night's observations.  Over the course of the observations, therefore,
the inclusion and exclusion of any host galaxy features within the
slit would have varied.  To combat this, a narrow aperture was used
during the extraction procedure, rather than performing an optimal
extraction using a weighted function over the entire cross-dispersive
axis.  To check whether any emission from extended regions had been
inadvertently included in the extraction aperture in spite of our precautions, a small ($\sim$20 pixels) section
of the cross-dispersive axis covering the [O {\sc iii}] 5007{\AA} line
of each 2-dimensional spectrum was summed and then checked to see whether there had been any
contamination by nearby features such as H {\sc ii} regions.  No
variable contamination was found.

\begin{figure}
	\centerline{\psfig{figure=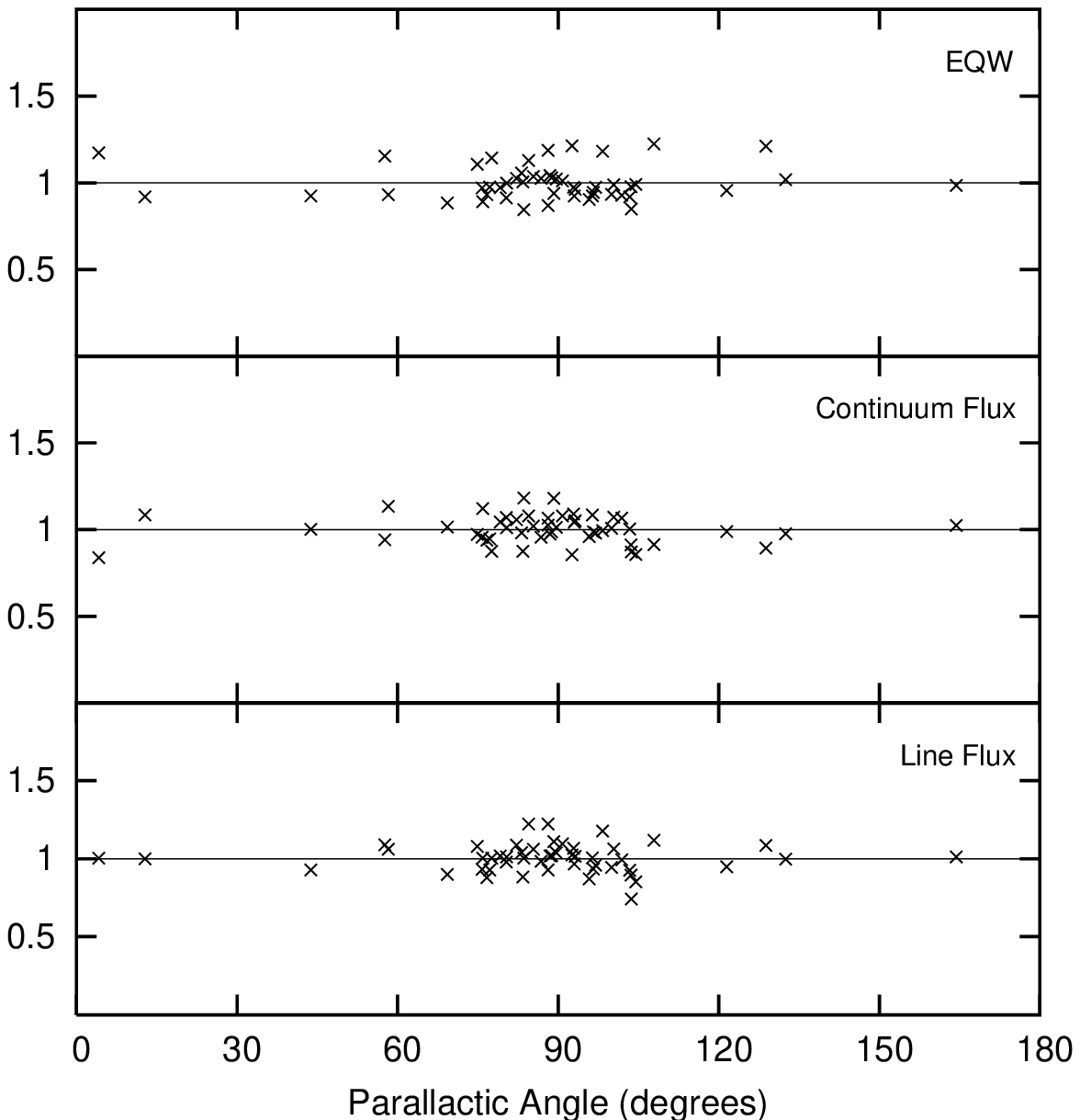,width=8.0cm}}
	\caption[]{Normalised equivalent width (top panel), continuum flux
(middle panel) and line flux (bottom panel) versus parallactic angle
for the [O {\sc iii}] 5007 line as measured from the 2 arcsec slit data.}
	\label{pa_vs_eqw}
	\end{figure}

To double-check this, the equivalent widths, line fluxes and continuum
fluxes were plotted versus the slit PA to check for correlations
(Figure \ref{pa_vs_eqw}).  No correlations are seen with any of these
variables, and so we are confident that any observed variability does
not arise from variable contributions from H {\sc ii} or star-forming
regions.  The lack of correlations \emph{also} rules out variability
resulting from changing ellipticity of the host galaxy core in the
slit.

Poor centring either of the slit on the galaxy or of the extraction
aperture during the data reduction could lead to variations in flux
density.  The use of two slits was intended primarily to take the
former point into account: if the observed variability in the 2
arcsec slit data was due solely to mis-centring of the galaxy in the
slit, we would not expect this to be as significant a problem in the
wider 8 arcsec slit data.  Since the variability has similar shape and
amplitude for both slits, we feel that we can confidently rule
this out as the source of variability.  Also, as mentioned above, the
extraction aperture used during the data reduction was checked and no
evidence that the aperture failed to collect all the light from the
Seyfert nucleus or that the host galaxy contribution varied
significantly was found.

Next, we checked whether any variations could be due to changes in
seeing conditions during the observations.  Three possibilities are
considered here:
(1) the variations are entirely due to seeing changes,
(2) the source has some small intrinsic variability, but this is
exaggerated by changes in flux brought about by changes in seeing, and
(3) the relative sizes of the NLR and continuum-emitting regions change
depending on the seeing conditions.  The first possibility can again
be discounted by looking at the variations seen in both the 2 arcsec
and 8 arcsec slit data.  If the variability is entirely due to seeing changes,
one would expect that any variations seen in the 2 arcsec slit data
would get `washed out' and be of considerably lower amplitude in the 8 arcsec
slit data because even with poor seeing all of the nuclear light
should pass through the wider slit.  Also, since the equivalent width is a measure of
the ratio between line and continuum flux, if all the variations in
the line and continuum fluxes are entirely artificial and from the
same source, we would expect to see \emph{no}
changes in the equivalent widths.  In fact the amplitude of
variability in both data sets is similar and so changes in equivalent
width cannot be due entirely to changes in seeing. 

\begin{figure}
	\centerline{\psfig{figure=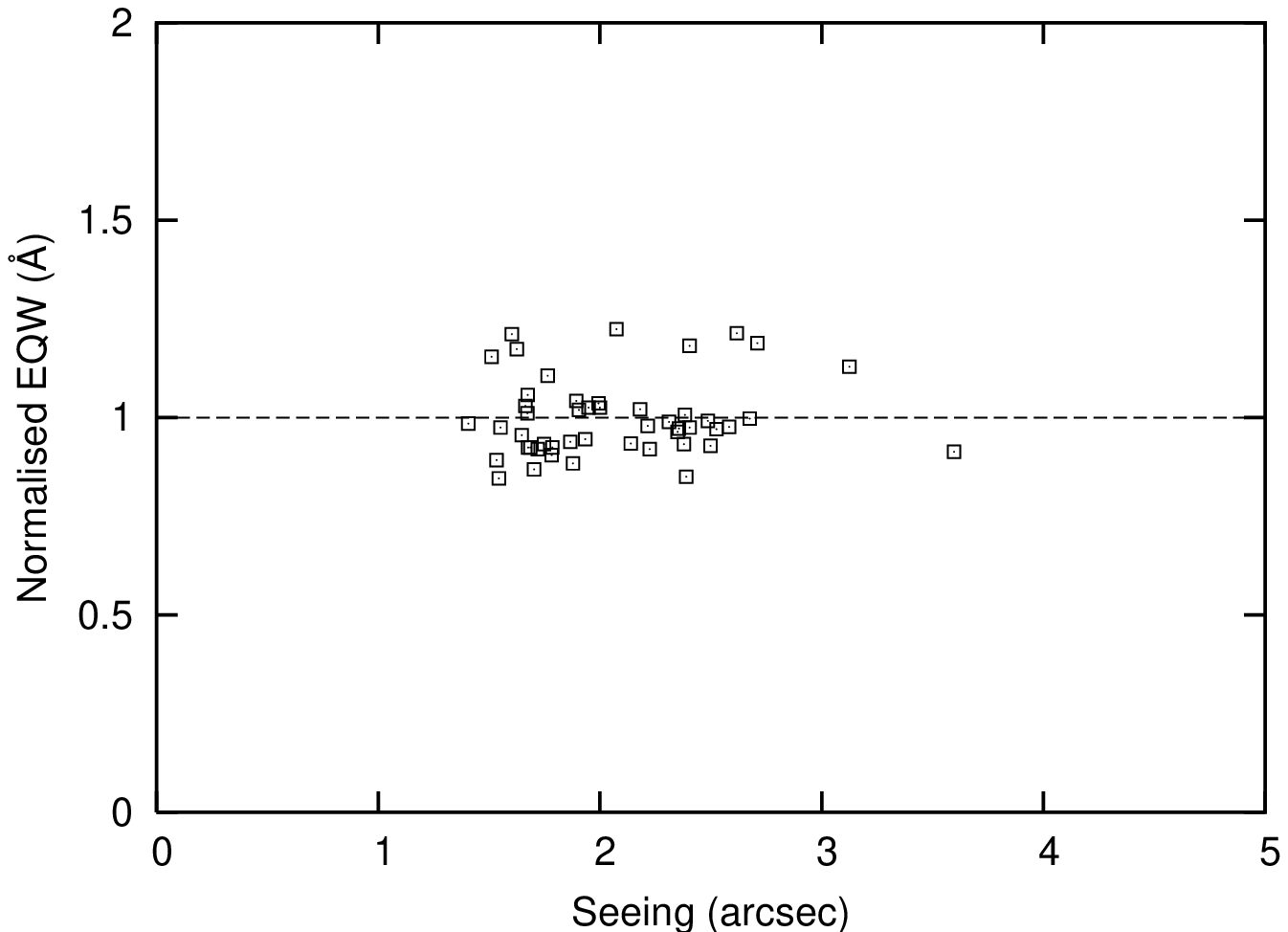,width=8.0cm}}
	\caption[]{Normalised equivalent width versus the FWHM of the
seeing disc.  The EQW and FWHM measurements were made using the [O
{\sc iii}] 5007 line from spectra taken using the 2 arcsec slit
width.}
	\label{seeing_vs_eqw}
	\end{figure}

It should be possible to determine whether any intrinsic source variations are
exacerbated by variations due to poor seeing by investigating whether
there is a correlation between the size of the seeing disc and the
equivalent width measurement, since it might reasonably be expected
that the magnitude of any effect would be directly related to the
quality of the seeing at the time.  To measure the size of the seeing
disc directly, a narrow section of each NGC 4395 spectrum close to the
[O {\sc iii}] 5007 emission line was extracted along the dispersion
axis, and a singe Gaussian fit to the resulting cross-dispersive
profile.  The FWHM of this fit was then used as a direct measurement
of the seeing throughout each night.  The equivalent width measurements
for the [O {\sc iii}] 5007 line are shown plotted versus FWHM in Figure \ref{seeing_vs_eqw}.  No correlation between the
parameters is seen.  Since any effect should be most readily observed
in the 2 arcsec slit data (shown here), we discount this as the source of the NGC
4395 variability.


\begin{figure}
	\centerline{\psfig{figure=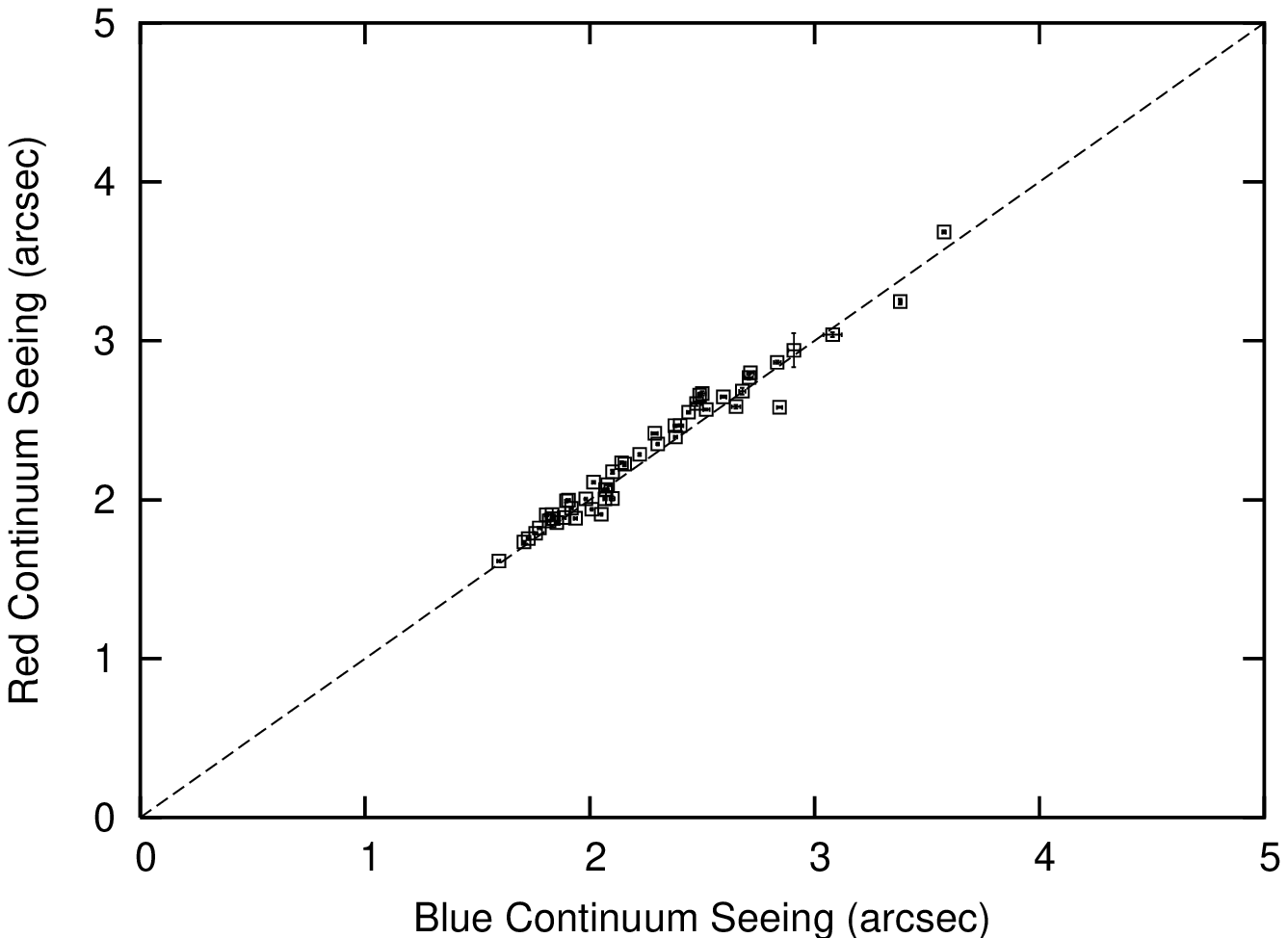,width=8.0cm}}
	\caption[]{FWHM of the cross-dispersive profiles measured for
the red continuum region against that of the blue continuum region,
from the 2 arcsec slit spectra.  Also shown is the 1:1 line for comparison.}
	\label{cont_reg_seeing}
	\end{figure}

\begin{figure}
	\centerline{\psfig{figure=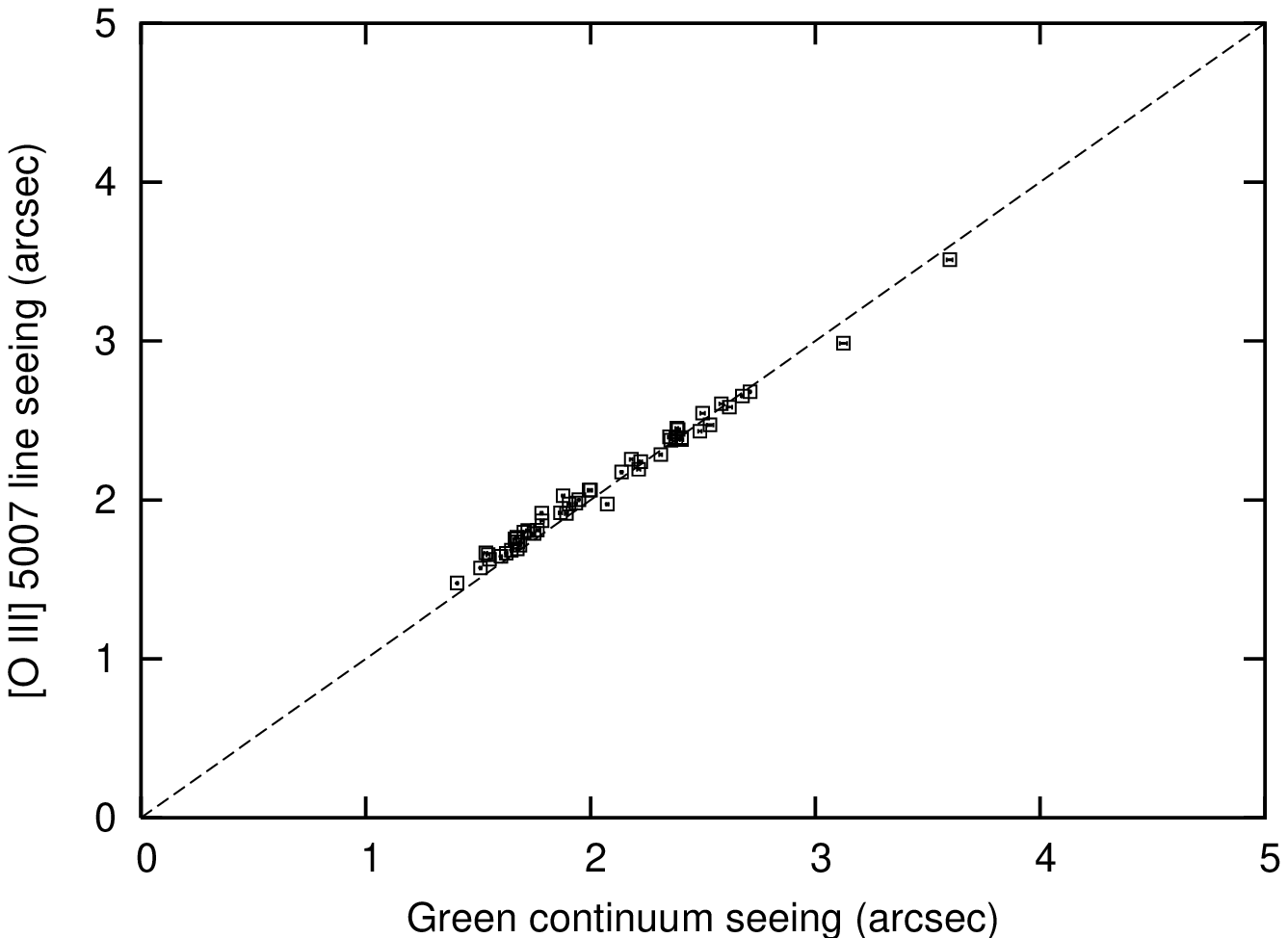,width=8.0cm}}
	\caption[]{FWHM of the cross-dispersive profiles measured for
the [O {\sc iii}] 5007 line and nearby continuum region.  Again the
1:1 line is shown for comparison.}
	\label{linevcont_seeing}
	\end{figure}

In addition to the [O {\sc iii}] 5007 line, continuum sections close
to the [O {\sc ii}] 3727 and [O {\sc i}] 6300 lines were extracted and
similar Gaussian fits applied to the cross-dispersive profiles.  The
same procedure was also applied to the spectral regions containing the
lines themselves.  Since the AGN core will be a blue point source
whereas any host galaxy contamination will be from diffuse and relatively red
starlight, comparing the FWHM of the blue and red continuum regions
checks whether changes in the quality of the seeing affect the seeing
disc for emission arising from different parts of the nuclear region in
different ways.  The FWHM for the red and blue continuum regions are
shown plotted against one another in Figure \ref{cont_reg_seeing},
together with the 1:1 line for comparison.  Similarly, changes in
seeing could affect emission from the continuum-emitting source
differently to emission from the NLR, which could be partially
resolved under good conditions (although as mentioned in Section 2
this is unlikely).  To test this, the FWHM for the extracted regions
containing the [O {\sc iii}] 5007 emission line and nearby continuum
region are shown plotted in Figure \ref{linevcont_seeing}, again with the 1:1 comparison
line.  There is no evidence in either Figure \ref{cont_reg_seeing} or
\ref{linevcont_seeing} for significant deviations from the 1:1 correlation, so from this we
conclude that the seeing changes have not had a significant
differential effect on emission from different regions and hence this
unlikely to be the source of the variability.  From Figure \ref{linevcont_seeing} we also
conclude that the NLR is not resolved in any of the observations and
so this is not likely to cause spurious variability.


The last remaining likely systematic error is atmospheric transparency
variations.  As discussed in Section 3.1, although we used standard
star observations to correct trends, short time-scale variations could
remain at the level of a few per cent.  In the next section, we show
that the scatter in measured line fluxes is consistent with this
effect, whereas continuum fluxes show a larger scatter.  Overall
then, we have considerable confidence that the continuum variations
show real intrinsic variability of the nuclear source.

\subsection{Variability Analysis}

\begin{figure}
	\centerline{\psfig{figure=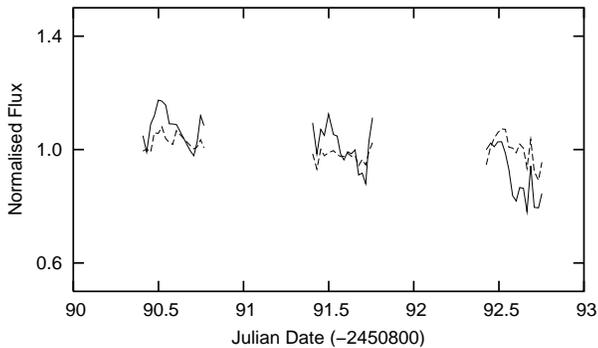,width=8.0cm}}
	\caption[]{Mean-normalised [{\sc oiii}] 5007 line flux (dashed
line) and continuum flux (solid line) for the 8 arcsecond slit.}
	\label{contandline}
	\end{figure}

\begin{figure}
	\centerline{\psfig{figure=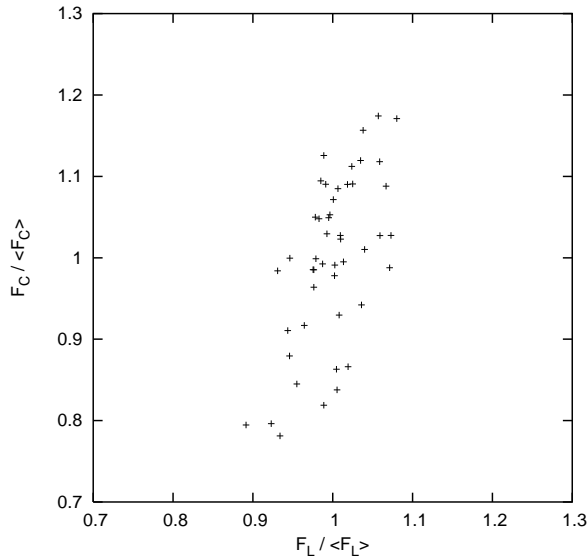,width=8.0cm}}
	\caption[]{Normalised continuum flux versus normalised narrow
line flux for the [{\sc oiii}] 5007 line measured from the 8 arcsec
slit data.}
	\label{8_cont_vs_narrowline}
	\end{figure}

Figure \ref{contandline} shows the mean-normalised line and continuum
fluxes for the [O {\sc iii}] 5007 line measured from the 8 arcsec slit
data.  The absolute flux calibration accuracy achieved for the wider
slit is $\sim$4 per cent
for all 3 nights, based on the deviation of the [O {\sc iii}] 5007
line fluxes from the absolute flux calibration.  The accuracy for the 2 arcsec slit data is somewhat
lower because these data suffered from additional fluctuations
attributed to changes in seeing across each night.  The mean absolute line
and continuum fluxes and fractional variation (defined as the ratio of
the standard deviation and the mean flux) for the three narrow
lines for which the equivalent widths were determined are given in
Table \ref{variability_pars}.  The figures for the 8 arcsec data
clearly show larger variances in the continuum fluxes than in the line
fluxes, suggesting intrinsic source variability.  Figure
\ref{8_cont_vs_narrowline} shows the continuum and line fluxes plotted against one another.  This not only shows
the greater continuum variability, but also shows a correlation
between the absolute flux measurements that suggests some small
uncorrected systematic errors remain.  This suggests the EQW will give
the most accurate determination of intrinsic variability.


	\begin{table*}
	\caption{Line and continuum variability parameters.}
	\begin{center}
	\begin{tabular}{lccccc@{\hspace{11mm}}cccc}

\hline
\hline

Band &  Night & \multicolumn{4}{c}{2 arcsec slit} &
\multicolumn{4}{c}{8 arcsec slit} \\
\AA &  & $\bar{F}_l\,^a$ & $\bar{F}_c\,^b$ & $\sigma_l$ / $\bar{F}_l$ &
$\sigma_c$ / $\bar{F}_c$ & $\bar{F}_l\,^a$ & $\bar{F}_c\,^b$ & $\sigma_l$ / $\bar{F}_l$ & $\sigma_c$ / $\bar{F}_c$  \\

\hline
$[{\rm \scriptstyle{OII}}]$ 3727 & 1 & 5.429 & 9.040 & 0.073 & 0.096 &
6.515 & 9.500 & 0.061 & 0.088 \\
 & 2 & 4.091 & 8.687 & 0.086 & 0.087 & 6.106 & 9.285 & 0.049 & 0.093 \\
 & 3 & 5.611 & 8.228 & 0.136 & 0.104 & 6.132 & 7.943 & 0.067 & 0.133 \\
 & All & 5.297 & 8.667 & 0.114 & 0.103 & 6.239 & 8.913 & 0.067 & 0.129
\\
\\
$[{\rm \scriptstyle{OIII}}]$ 5007 & 1 & 26.54 & 5.820 & 0.074 & 0.085 & 28.26 & 6.866 & 0.027 & 0.056 \\
 & 2 & 25.13 & 5.638 & 0.077 & 0.058 & 26.93 & 6.423 & 0.022 & 0.068 \\
 & 3 & 28.79 & 5.516 & 0.070 & 0.085 & 27.47 & 5.782 & 0.053 & 0.100 \\
 & All & 26.72 & 5.670 & 0.091 & 0.080 & 27.53 & 6.348 & 0.041 & 0.102 \\
\\
$[{\rm \scriptstyle{OI}}]$ 6300 & 1 & 2.812 & 4.721 & 0.093 & 0.097 & 3.125 & 5.861 & 0.033 & 0.058 \\
 & 2 & 2.692 & 4.521 & 0.070 & 0.052 & 2.926 & 5.497 & 0.028 & 0.050 \\
 & 3 & 2.977 & 4.519 & 0.078 & 0.089 & 2.785 & 4.686 & 0.046 & 0.096 \\
 & All & 2.820 & 4.595 & 0.091 & 0.085 & 2.941 & 5.341 & 0.059 & 0.113
\\
\hline
\hline
\multicolumn{10}{l}{$^a$Line fluxes in units of $10^{-14}$ ergs s$^{-1}$ cm$^{-2}$}\\
\multicolumn{10}{l}{$^b$Continuum fluxes in units of $10^{-16}$ ergs
s$^{-1}$ cm$^{-2}${\AA}$^{-1}$}
	\end{tabular}
	\end{center}
	\label{variability_pars}
	\end{table*}
		

The optical light curves constructed using the EQW measurements in
Tables \ref{flux_meas_2} and \ref{flux_meas_8} are shown in Figures \ref{lc_2_all_scaled} and \ref{lc_8_all_scaled} respectively, while the
fractional variation in EQW for the three narrow lines and both slits
are given in Table \ref{eqw_pars}.  The average sampling interval was
31.6 min and 33.0 min for the 2 arcsec and 8 arcsec slit data respectively, and
the large gaps in the light curve represent daylight periods.  The light curves show that the nucleus in NGC 4395 varied
significantly during the observing campaign, including over short ($<$ 8 hr)
time-scales, with the changes similar in shape for all three lines and
for both slit widths.
The largest variability occurs on night 3, with a $\sim$
50 per cent increase in equivalent width (at [O {\sc ii}] 3727{\AA})
occurring over a time-scale of 5 hours, corresponding to a
\emph{decrease} in continuum luminosity of 35 per cent.  Similar variability, but with smaller amplitude, is also seen
in the green and red continuum regions.  Small amplitude ($\sim$ few
per cent) variability is seen over time-scales as short as the sampling
interval, but given the measurement errors this is likely to be of only marginal
significance.

\begin{figure}
	\centerline{\psfig{figure=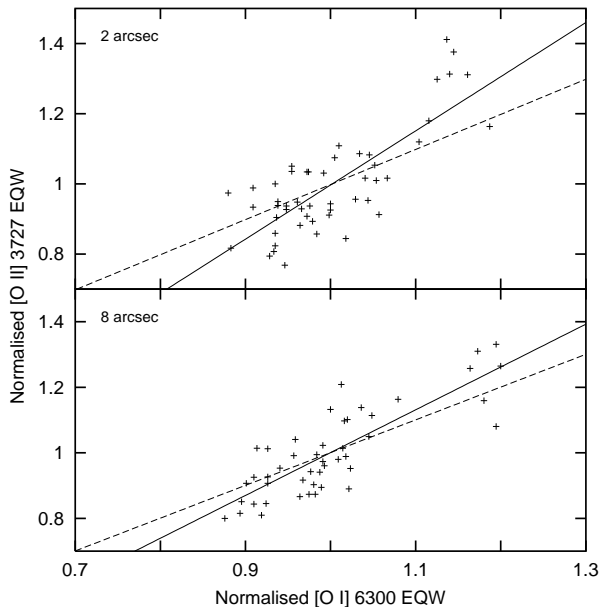,width=8.0cm}}
	\caption[]{[{\sc o ii}] 3727 normalised equivalent width versus
[{\sc o i}] 6300 normalised equivalent width for both the 2 arcsec slit
(top) and 8 arcsecond slit (bottom).  In both cases the solid line
represents the best-fitting straight line through the points and the
dashed line represents the 1:1 gradient correlation.  The gradients of the
best-fit lines are 1.54 and 1.33 for the 2 arcsec and 8 arcsec slit
data respectively.}
	\label{3727_vs_6300}
	\end{figure}

	\begin{table}
	\caption{Fractional variation in the EQW measurements for the
three narrow lines.}
	\begin{center}
	\begin{tabular}{lccc}

\hline
\hline
Band &  Night & 2 arcsec slit & 8 arcsec slit\\
\AA & & $\sigma_{\scriptscriptstyle{EQW}}/\bar{EQW}$ &
$\sigma_{\scriptscriptstyle{EQW}}/\bar{EQW}$\\
\hline
$[{\rm \scriptstyle{O\; II}}]$ 3727  & 1 & 0.090 & 0.091 \\
 & 2 & 0.081 & 0.104 \\
 & 3 & 0.162 & 0.133 \\
 & All & 0.147 & 0.135 \\
\\
$[{\rm \scriptstyle{O\; III}}]$ 5007  & 1 & 0.053 & 0.043 \\
 & 2 & 0.053 &  0.054 \\
 & 3 & 0.090 & 0.070 \\
 & All & 0.099 & 0.089 \\
\\
$[{\rm \scriptstyle{O\; I}}]$ 6300  & 1 & 0.040 & 0.050 \\
 & 2 & 0.050 & 0.047 \\
 & 3 & 0.077 & 0.081 \\
 & All & 0.075 & 0.083 \\
\hline
\hline
	\end{tabular}
	\end{center}
	\label{eqw_pars}
	\end{table}
To test whether there was any difference in the variability with
colour, the EQW measurements for the [O {\sc ii}] 3727 and [O {\sc i}]
6300 line were plotted against each other, as shown in Figure
\ref{3727_vs_6300}.  A least-squares fit was then performed in the
usual way to determine the best-fitting straight line.  The best-fit
solutions have gradients of 1.54 and 1.33 for the 2 arcsec and 8
arcsec data respectively, although a 1:1 relation gave a poorer fit
(reduced $\chi^2=5.92$ for the 2 arcsec slit data) but could not
entirely be ruled out. This suggests that there was greater
variability in the [O {\sc ii}] 3727 EQW than the [O {\sc i}] 6300
EQW, which is consistent with the findings of Clavel et al. (1991) that the spectra of AGN harden as they
brighten.  The effect could potentially be caused by starlight
contamination, however the surface brightness of the host galaxy bulge is
low, arguing against a large degree of contamination.  The difference
between the two lines is larger for the 2 arcsec data than the 8
arcsec data, which is consistent with the 8 arcsec data being affected
more by contamination from the host galaxy starlight, suggesting a
small but real residual colour effect. 


\begin{figure}
	\centerline{\psfig{figure=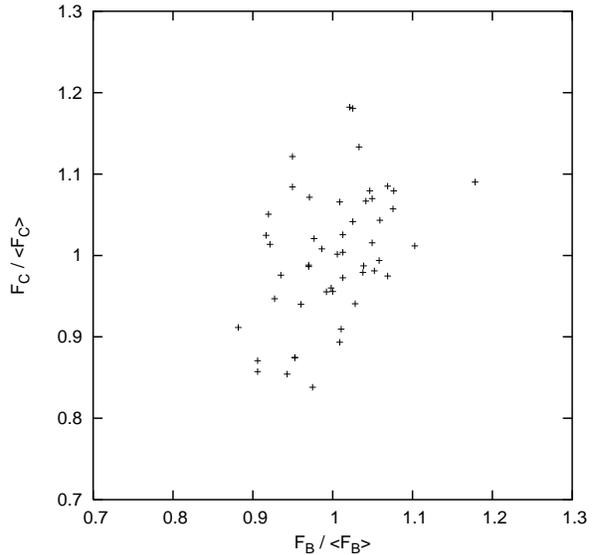,width=8.0cm}}
	\caption[]{Normalised continuum flux measured under the [{\sc
oiii}] 5007 line (using a low-order polynomial fit to the continuum) versus the
normalised flux measured for the broad component of the H$\beta$ line.
Measured using the 2 arcsec slit data.} 
	\label{2_cont_vs_broadline}
	\end{figure}

\begin{figure}
	\centerline{\psfig{figure=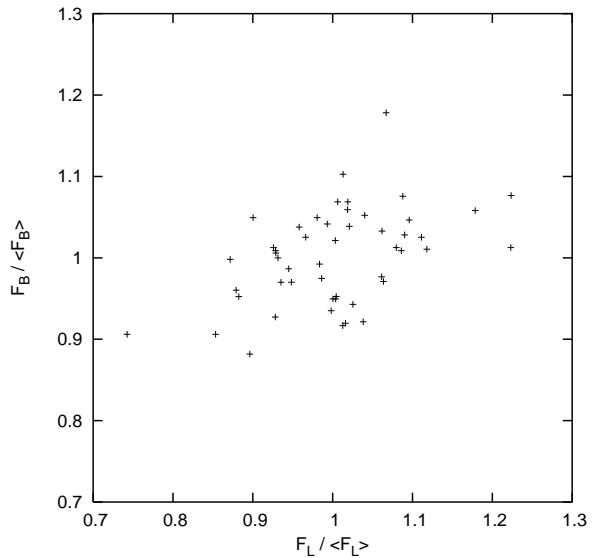,width=8.0cm}}
	\caption[]{Normalised flux for the broad component of the
H$\beta$ line versus line flux measured for the [{\sc oiii}] 5007
line.  Measured using the 2 arcsec slit data.} 
	\label{2_broad_v_narrow}
	\end{figure}

No significant variability in the broad H$\beta$ flux was observed.
If the variability in NGC 4395 is qualitatively the same as that in
other Seyfert 1s such as NGC 5548 (see Section 5), we would expect to
see roughly the same variations in the H$\beta$ line as the nearby
continuum, with a lag induced by the light travel time between the
accretion disc and broad-line region (Peterson et al. 1992). However, the signal-to-noise
ratio of the broad component is poor and proved difficult to fit a
Gaussian profile to, so it is difficult to make quantitative statements.
Figure \ref{2_cont_vs_broadline} shows the mean-normalised continuum
flux measured for the [O {\sc iii}] 5007 line plotted against that for
the H$\beta$ broad component.  The scatter is greater than that seen
in Figure \ref{8_cont_vs_narrowline}, but there is still greater variance in the continuum
flux than the broad H$\beta$ flux.  Figure \ref{2_broad_v_narrow}
shows the mean-normalised broad H$\beta$ flux against the
mean-normalised flux for the narrow component.  The scatter is
considerable for both parameters - the fractional variation over all 3
nights for the broad and narrow components is 5.9 and 9.5 per cent
respectively.  The apparent correlation is expected and is a result of
allowing the amplitude of both fitted Gaussians to vary as free parameters.


%

\section{Discussion}

\subsection{Comparison with NGC 5548}

The short time-scale optical variability seen in NGC 4395 (shown in Figures
\ref{lc_2_all_scaled} and \ref{lc_8_all_scaled}) is extremely rare in Seyfert
galaxies.  This suggests that NGC 4395 does indeed have a small black
hole rather than an extremely low accretion rate.  Simultaneous X-ray and optical observations of another
low-luminosity Seyfert 1, NGC 4051 (Done et al. 1990) found no evidence
for significant optical variability during a 6-day period in which the
X-ray flux varied by up to a factor of 2.  The total variability in
the B-band over the entire observing period was just 1.4 per cent, with the
optical flux remaining constant to within 1 per cent during the periods of
highest X-ray variability.

The variability of both continuum and emission lines in the Seyfert 1
galaxy NGC 5548 has been well-studied at both UV and optical
wavelengths in order to determine the structure of the
unresolved emission line regions.  NGC 5548 was monitored in
the UV every 4 days for a total of 8 months using \emph{IUE} (Clavel
et al. 1991) and every few days over a period of 10 months using a
number of telescopes in the optical (Peterson et al. 1991).  The
large quantity of variability data at these wavelengths makes it an ideal
object to use as a comparison with NGC 4395.

To compare NGC 4395 and NGC 5548, the variability power density
spectrum (PDS) for NGC 5548 for the UV light curve was obtained from
Krolik et al. (1991).  The stochastic nature of the variability
in AGN means that the PDS is the best way to probe the distribution of
variability amplitude with time-scale.  Given that the
effect of red-noise means that the variability amplitude depends on
the observation length, the PDS slope must be known before the
fractional variability of two objects can be compared.  Here we assume
that the slopes of the PDS for NGC 5548 and NGC 4395 are the same and
that the difference in black hole mass changes the frequency scaling,
resulting in a change in the PDS normalisation at a given frequency.  This is consistent with the recent results of Vaughan et al. (2004), who
suggest that the large X-ray variability amplitudes seen in both NGC 4051
and NGC 4395 could be the result of the power spectra in these objects
being shifted to higher frequencies in comparison with AGN with more
massive black holes.  

A
power-law of the form $P(\nu)=K\nu^{-\alpha}$ was fitted to the NGC
5548 PDS.  The final few points are noisy due to the 4-day sampling
interval and so these were excluded from the fit.  The best-fit
solution gave a power-law index of $\alpha = 2.8$, which was used to
extend the PDS to the higher frequencies sampled by the NGC 4395 light
curves discussed here.  Since the UV fluctuations for NGC 5548 have
larger amplitude than the optical fluctuations, the normalisation of
the UV PDS was scaled using the ratio between the fractional
variations for the UV and optical light curves calculated using the
data from Clavel et al. (1991) and Peterson et al. (1991)
respectively.  To avoid the effects of red-noise on the variances,
data sets of the same length were used for this. 

The scaled PDS was used to calculate the expected variances in the
optical light curves for NGC 5548 over time-scales of 8 and 72 hours.
The expected fractional variations over these time-scales were
calculated to be 0.04 per cent and 0.3 per cent respectively.  The process was
repeated, this time using the original UV PDS and the expected
fractional variations in the UV for NGC 5548 were calculated to be
0.09 per cent (8 hr) and 0.7 per cent (72 hr).  Neither the expected optical or UV
fractional variability for NGC 5548 is consistent with the 6-13 per cent
variation observed in the continuum flux for NGC 4395.  This analysis
confirms that NGC 4395 has larger short time-scale variability than
other Seyfert galaxies.  A similar result holds in X-rays (Iwasawa
et al. 2000; Shih et al. 2003).  It seems plausible that this is because
the black hole is smaller and so time-scales are shorter.  In the next section we address this question
quantitatively. 

\subsection{Does variability scale with black hole mass?}

The precise relationship between variability and black hole mass is
dependent upon the precise origin of the variability.  If the
time-scales for variability are dictated principally by the light
travel time across the accretion disc, then one might expect a direct
scaling between variability time-scale and black hole mass.
Considering a more detailed model in which the time-scale for
variability at different wavelengths is dictated by the sound-crossing
time-scale gives a different answer; this is discussed below.

To investigate direct scaling between black hole mass and variability
we assume that the {\it slope} of the variability power spectra for
both NGC 4395 and NGC 5548 are the same, and that the different black
hole masses result in a simple translation of the power spectrum along
the frequency axis.  Then the ratio of black hole mass can be used to
predict the variability that we would see in NGC 4395 using the NGC
5548 variability for which the power spectrum is already known.  We
use a black hole masses of $10^5{\mdot}$
(Kraemer et al. 1999; Shih et al. 2003) and
$6.1 \times 10^7{\mdot}$ (Wandel, Peterson \& Malkan 1999) for NGC
4395 and NGC 5548 respectively.  This implies that a time-scale of 200
days in NGC 5548 should correspond to a time-scale of 8 hours in NGC
4395.  Using the UV power spectrum for NGC 5548 we find that
variability of 29 per cent is expected over this time-scale.  This is much larger than the
variability seen for NGC 4395 over an 8-hour period for any optical
wavelength.  If however we consider the optical variability of NGC 5548 over
200 days, we expect variability of order 12 per cent, which is much closer to
that seen in NGC 4395.  It is unclear at present whether a
direct scaling of the two power spectra is valid.  Consequently we
have also investigated a more detailed model in which we have
attempted to take some of the properties of the accretion disc into account.

We have modelled the central engine of NGC 4395 using a simple
accretion disc for which the emitted spectrum is a sum of blackbodies.  Considering the release of thermal binding energy at a radius R gives
a blackbody effective temperature of 
\begin{equation}
T(R) = \left(\frac{GM_{\bh}\dot{m}}{8\pi R^3\sigma}\right)^4
\end{equation}
where $M_{\bh}$ is the mass of the central black
hole, $\dot{m}$ is the accretion rate, $\sigma$ is the
Stefan-Boltzmann constant and all other symbols take their usual
values.  Obviously the precise formula for $T(R)$ depends on the specific
model adopted, but the binding energy formula should be good enough
for our approximate scaling calculation below.  If we assume that the nucleus is time-steady and has
an accretion rate $\dot{m}=(6.32 M_{\bh}/{\eta}c^2)(L/L_{\rm{\scriptscriptstyle{E}}})$ (in SI units), where
$L_{\rm{\scriptscriptstyle{E}}}$ is the Eddington luminosity, the above temperature relation then gives
\begin{equation}
R^3 \propto \frac{M_{\bh}^2}{T^4}
\left(\frac{L}{L_{\rm{\scriptscriptstyle{E}}}}\right)
\label{r-m-reln}
\end{equation}

If the size of the emitting region and the variability time-scale scales
with black hole mass, then it should be possible to calculate the
scaling factor for NGC 4395 and NGC 5548.  We use an accretion rate of
$0.0017L_{\rm{\scriptscriptstyle{E}}}$ (Lira et al. 1999) for NGC 4395, and
an accretion rate of 0.015$L_{\rm{\scriptscriptstyle{E}}}$ for NGC 5548
(Wandel et al. 1999).  From (\ref{r-m-reln}) we find that
for a fixed wavelength (and hence temperature) the time-scales for
variability for the two objects should scale by a factor $(6.1 \times
10^7/10^5)^{2/3}(0.015/0.0017)^{1/3} = 149$.  So a time-scale of 8
hours in NGC 4395 (i.e. one observing night) should correspond to
about 50 nights for NGC 5548.  A randomly-selected sample of 10
sections of 50 days from the optical flux values given in Table 9 of
Peterson et al. (1991) gives an average fractional deviation of
~10.9 per cent.  This value agrees well with the average fractional deviation
of  10.0 per cent in the [O {\sc ii}] 3727 EQW measurements for NGC 4395 over
time-scales of 1 night.  This shows that the anomalous variability of
NGC 4395 is quantitatively consistent with that expected for a small
central black hole. 

\subsection{Absolute variability time-scales}

Above we showed that the relative variability of NGC 4395 is
consistent with simple accretion disc models.  However there are
problems when we consider the absolute variability time-scales.  Considering the light- and sound-crossing time-scales for a particular
black hole mass gives lower and upper limits on the time-scales
for variability using $R \propto M_{\bh}^{2/3}
\propto \Delta t$.  If the mass of the central engine in NGC 4395 is
of the order $10^5$M$_{\odot}$ and the temperature of order $10^4$K, then using a value of $10^4$ ms$^{-1}$
for the sound speed gives a
light-crossing time-scale of 28 min and sound-crossing time-scale of 1.6
years for the time-steady, Eddington-limited case if the optical
emission is emitted at $1720R_S$ .  A more realistic estimate (Lira
et al. 1999)
puts the accretion rate at $\sim$10$^{-3}L_{\rm{\scriptscriptstyle{E}}}$, giving light- and
sound-crossing time-scales of 2.8 min and 59 days respectively.
Instabilities in the disc would be expected to propagate at the sound
speed, and therefore we would not expect to see variability on
time-scales shorter than this.  This shows that variability time-scale
is therefore a problem for NGC 4395 as it is for other Seyfert
galaxies. 

\section{Summary and Conclusions}

We present optical spectroscopic data for the least-luminous Seyfert
1, NGC 4395, covering three nights during which the nucleus was
observed approximately every 30 minutes.  Two slit widths were used,
enabling us to minimize or quantify systematic errors and aperture
effects, giving an absolute flux calibration accuracy of 5-10 per cent.  To
reduce the effect of any remaining systematic errors, equivalent width
measurements of three narrow lines in the blue, green and red regions
of the spectrum were used to quantify the continuum variability, while
absolute measurements of the broad-component flux of the H$\beta$ line
were used to constrain variability in the broad lines.

The continuum in NGC 4395 was variable over all three nights, with the
greatest amplitude of variability occurring during night 3.  The
variability appeared to be simultaneous for all three continuum
regions, with greater variability seen in the continuum flux under the
[O {\sc ii}] 3727 line
than the continuum under the [O {\sc i}] 6300 line.  The observed hardening of the
spectrum with increasing luminosity is consistent with the findings of
Clavel et al. (1991), although contamination by starlight cannot be
completely ruled out.  No measurable variability was detected in the
broad H$\beta$ flux.

The nucleus in NGC 4395 was compared with another well studied Seyfert
1, NGC 5548.  NGC 4395 was found to vary with greater amplitude over
shorter time-scales than NGC 5548.  A simple accretion disk model was
used to calculate the expected scaling between NGC 4395 and NGC 5548
if the time-scale for variability scaled with the mass of the central
black hole, and the results were consistent with the observed optical
variability for NGC 5548 if NGC 4395 has a black hole with a mass approximately
two orders of magnitude smaller than that for NGC 5548.  The observed variability was not consistent
with the expected absolute time-scales for variability if instabilities
propagate through the accretion disc at sound speed.

\section*{Acknowledgements}

JES acknowledges a PPARC research studentship.  The Isaac Newton
Telescope is operated on the island of La Palma by the Royal Greenwich
Observatory in the Spanish Observatorio del Roque de los Muchachos of
the Instituto de Astrofisica de Canarias. {\sc iraf} is distributed by
the National Optical Astronomy Observatories, which are operated by
AURA, Inc., under cooperative agreement with the National Science Foundation.

\label{lastpage}
\end{document}